\documentclass[fleqn,usenatbib]{mnras}
\hypersetup{linkcolor=blue,citecolor=blue,filecolor=cyan,urlcolor=magenta}

\usepackage{newtxtext,newtxmath}
\usepackage[T1]{fontenc}

\DeclareRobustCommand{\VAN}[3]{#2}
\let\VANthebibliography\thebibliography
\def\thebibliography{\DeclareRobustCommand{\VAN}[3]{##3}\VANthebibliography}

\usepackage{graphicx}	
\usepackage{amsmath}	
\usepackage{bm}

\title[AR Sco Particle Dynamics]{Towards Modelling AR Sco: Generalised Particle Dynamics and Strong Radiation-Reaction Regimes}

\author[L. Du Plessis et al.]{
L. Du Plessis,$^{1}$\thanks{E-mail: louisdp95@gmail.com}
C. Venter,$^{1,2}$
A.K. Harding$^{3}$
Z. Wadiasingh$^{4,5,6,1}$
C. Kalapotharakos$^{4}$
and P. Els$^{1}$
\\
$^{1}$Centre for Space Research, North-West University, Private Bag X6001, Potchefstroom 2520, South Africa\\
$^{2}$National Institute for Theoretical and Computational Sciences, South Africa\\
$^{3}$Theoretical Division, Los Alamos National Laboratory, Los Alamos, NM 58545, USA\\
$^{4}$Astrophysics Science Division, NASA Goddard Space Flight Center, Greenbelt, MD 20771, USA \\
$^{5}$Department of Astronomy, University of Maryland, College Park, MD 20742-4111, USA\\
$^{6}$Center for Research and Exploration in Space Science and Technology, NASA/GSFC, Greenbelt, Maryland 20771, USA  \\
}

\date{Accepted 2024 July 21. Received 2024 June 16; in original form 2024 February 29}

\pubyear{2024}

\begin{document}
\label{firstpage}
\pagerange{\pageref{firstpage}--\pageref{lastpage}}
\maketitle

\begin{abstract}
Numerical simulations of relativistic plasmas have become more feasible, popular, and crucial for various astrophysical sources with the availability of computational resources. The necessity for high-accuracy particle dynamics is especially highlighted in pulsar modelling due to the extreme associated electromagnetic fields and particle Lorentz factors. Including the radiation-reaction force in the particle dynamics adds even more complexity to the problem, but is crucial for such extreme astrophysical sources. We have also realised the need for such modelling concerning magnetic mirroring and particle injection models proposed for AR Sco, the first white dwarf pulsar. This paper demonstrates the benefits of using higher-order explicit numerical integrators with adaptive time step methods to solve the full particle dynamics with radiation-reaction forces included. We show that for standard test scenarios, namely various combinations of uniform $E$- and $B$-fields and a static dipole $B$-field, the schemes we use are equivalent to and in extreme field cases outperform standard symplectic integrators in accuracy. We show that the higher-order schemes have massive computational time improvements due to the adaptive time steps we implement, especially in non-uniform field scenarios and included radiation reaction where the particle gyro-radius rapidly changes. When balancing accuracy and computational time, we identified the adaptive Dormand-Prince eighth-order scheme to be ideal for our use cases. The schemes we use maintain accuracy and stability in describing the particle dynamics and we indicate how a charged particle enters radiation-reaction equilibrium and conforms to the analytic Aristotelian Electrodynamics expectations.
\end{abstract}




\section{Introduction} \label{sec:1}
In recent years, kinetic simulations have been crucial to further our understanding of pulsar electrodynamics. These simulations have been used to numerically and self-consistently solve the structure and geometry of a pulsar magnetosphere and calculate the radiative processes along the particles' trajectories. The major limiting factor for these simulations is the large-scale separation between the gyro-period compared to the stellar rotation period. It is unlikely, even with drastically increasing computational power, that the problem will be solved via brute-force computations in the near future. Particle-in-Cell (PIC) approaches cope by scaling up the gyro period to computationally realistic scales by lowering the electromagnetic field values, particle Lortenz factors $\gamma$, and radiation-reaction forces (RRF) by orders of magnitude \citep[e.g.,][]{Cerutti2013, Kalapotharakos2018, Brambilla2018, Cruz2023}. The problem is that these parameters are many orders of magnitude different than what is realised in real pulsars. Simply re-scaling the parameters and forces to higher values after the simulation is questionable, since there are different considerations (that may not have been taken into account) when one is in the RRF limit at high $\gamma$ and field strengths \citep{Petri2023}. 

PIC models start by solving the Maxwell equations self-consistently to calculate the fields and use phenomenological descriptions of the plasma injection needed to fill the magnetosphere. These phenomenological descriptions, and how radiation losses are treated/scaled, are where the various models differ. After obtaining the field values, the macro-particle trajectories are calculated with the equations of motion using the new field values. The new position and velocity information is then used to calculate the new field values. Summaries and critiques of the different pulsar PIC schemes can be found in \citet{Kalapotharakos2018, Cruz2023}. 

In PIC codes, the particle trajectory calculations are a computationally demanding aspect, since they solve the full particle gyration with included classical RRF. Another technique to deal with the problem of the large difference in gyro-period vs stellar period is using a gyro-centric approach for the particle trajectories. One may simply solve the equation of motion describing the gyro-centric curve, allowing the use of much larger time or spatial steps, since the particle gyration is not solved. Two examples are the models of \citet{Harding2021} and \citet{Gruzinov2012}, where the latter uses an Aristotelian Electrodynamics (AE) approach. We will discuss these models as well as the differences and advantages of using a gyro-centric approach vs.\ solving the full particle gyration and directly compare the two methods in a future work.

Our overarching goal is to develop models to produce emission maps and spectra for the `white dwarf' pulsar AR Sco, which samples an interesting regime of plasma parameters that require a rethinking of current approaches (see below). Our focus in this paper is to test different numerical schemes and select the most appropriate one for this source and pulsar sources, while we defer detailed calibration and actual model fitting to follow-up works (du Plessis et al., in prep.). We will initially neglect the self-consistent calculations of the fields, i.e., how the particles modify the fields. It is much easier to test accuracy and calibration of different schemes with analytical fields; we will include self-consistent field calculations after we are confident in our particle dynamics. Thus we initially deviate from a full PIC scheme.

AR Sco is a unique, close white dwarf (WD) binary system observed to exhibit pulsed, non-thermal emission, with its radiation spectrum peaking in the optical range \citep{Marsh2016}. The inferred spin period for the WD is $117~\rm{s}$ with an orbital period of $3.56~\rm{h}$. Highly linearly polarised optical emission with a $180^{\circ}$ polarisation angle swing has been observed \citep{Buckley2017}. The system exhibits neither an indication of an accretion disc \citep{Marsh2016}, nor of an accretion column \citep{Takata2018, Garnavich2019}. 
The WD spin-down rate has been constrained \citep{Stiller2018, Gaibor2020}, yielding an inferred surface $B$-field of $\sim  5\times 10^{8} \, \rm{G}$. However, observations by \citet{Garnavich2020} set an upper limit for the $B$-field $(B< 100 \, \rm{MG})$, since no Zeeman splitting was observed in the WD emission line spectrum. 

The high inferred $B$-field of this WD is currently under contention. Most authors suggest a high $B$-field, assuming that the emission is powered by the WD spin-down and performing spectral fitting by invoking a non-thermal synchrotron radiation (SR) spectrum \citep{Geng2016, Buckley2017, Katz2017, Takata2017, Potter2018b, DuPlessis2019}. A popular model amongst these is the magnetic mirror scenario by \citet{Takata2017} in which particles are injected at the companion and become trapped in the white dwarf magnetosphere and are mirrored between the magnetic poles as they encounter severe SR losses at the mirror points. This mirroring idea was invoked by \citet{Potter2018b} as well and is the first scenario we want to investigate with our modelling. However, \citet{Lyutikov2020} argued that such a large $B$-field would imply a mass transfer rate of $\dot{M} \sim 10^{-4} \, M_{\odot}yr^{-1}$ to spin up AR Sco to its current spin period, which is $10^{5}$ times larger than that seen in similar binaries. But this argument assumes mass transfer purely via the companion L1 point and neglects another popular idea that allows a large WD $B$-field, namely clumpy mass transfer, which has also been proposed for AE Aquarii \citep{Wynn1995, Wynn1995b}. A propeller/intermediate polar model for AR Sco has been proposed by \citet{Lyutikov2020} and adopted by \citet{Garnavich2020}, since this allows for a smaller $B$-field $(B\sim 5 \times 10^{5} \, \rm{G})$. This may address the large mass transfer rate, but does not address the lack of observational signatures of an accretion column or a propeller scenario. Also unaddressed in this model is a self-consistent broadband radiation scenario that could fit the non-thermal spectrum with such a low $B$-field. Given the detection of the first AR Sco sibling by \citet{Pelisoli2023}, that paper uses a similar propeller model, but there does seem to be flaring observed in the sibling system \citep{Pelisoli2023b}. Interestingly, a magnetic WD evolution model was proposed by \citet{Schreiber2021} that may explain the high spin rate of AR Sco. This model suggests that the WD in a cataclysmic variable starts with no $B$-field and obtains a $B$-field via a rotational and crystallisation dynamo. Most of these models and others have been summarised in \citet{DuPlessis2022}.  
Given the large $B$-field and presence of mildly relativistic particles in this system, it is evident that more sophisticated modelling is required for AR Sco. No emission model has yet adequately modelled the light curves, spectra, and polarisation properties of AR Sco without the limiting assumption of super-relativistic particles with small pitch angles. The assumption of small pitch angles is inappropriate when magnetic mirroring (and strong cooling) of electrons could occur, as expected near the magnetic poles of the WD in AR Sco. Thus we have developed an emission model using fully general particle dynamics that can capture both the relativistic and non-relativistic motion of particles in large fields, with RRF. This is required to accurately capture the physics in AR Sco. We can also compare our methods to PIC approaches that incorporate RRF, like that of \citet{Cerutti2013}. Solving the general equations of motion is important to understand these systems where particle mirroring, focusing, and drift effects impact the particle trajectories and pitch angle evolution. Our study also serves as a validation for the gyro-centric approach and assesses the applicability of the latter. 

Solving the general equations of motion for pulsars or pulsar-like sources requires high computational accuracy. This is due to the presence of highly relativistic particles, large $B$- and $E$-fields, rapidly changing variables, and added RRF where the particles are in or close to the radiation-reaction limit. These calculations are extremely computationally demanding, thus we will investigate higher-order particle integrators with adaptive time steps to increase and balance accuracy and computational runtime. 

In this first study, we discuss how we solved our particle dynamics while including the RRF by showing accuracy and benchmark results for known test scenarios. We thus build up the complexity of the test scenarios to identify sufficiently accurate numerical schemes and techniques that can reliably solve the particle dynamics under the extreme conditions found in pulsars and pulsar-like systems. In our next work, we will present our calibration results between our code and those of the pulsar emission model of \citet{Harding2021}, given the aspects of overlap between AR Sco and canonical or millisecond pulsars; we will furthermore discuss our comparison to the Aristotelian Electrodynamics (AE) results of \citet{Gruzinov2012}. We will thus be using retarded dipole and force-free fields, and investigate different radiation mechanisms to generate skymaps and spectra for pulsar cases. Finally, in the third paper, we will present our light curve and spectral results for AR~Sco. 

In this paper, we start by discussing our methods for solving our particle dynamics in detail and considered test scenarios in Section~\ref{sec:2}. We will then present and discuss our results in Section~\ref{sec:3}, and finally we will make closing remarks in Section~\ref{sec:4}. 

\section{Method}\label{sec:2} 
In this section, we will discuss how we solve the particle dynamics that we will use in our upcoming pulsar simulations, with and without radiative losses, as well as the benchmarks and accuracy tests that we implemented.

To find a particle's momentum and position at the next given time step, we solve the relativistic Lorentz force formula given in cgs units:
\begin{equation}\label{Lorentz}				
\frac{d\mathbf{p}}{dt} = e\left( \mathbf{E} + \frac{c\mathbf{p}\times\mathbf{B}}{\sqrt{m^{2}c^{4}+\mathbf{p}^{2}c^{2}}}\right),
\end{equation}
where $\mathbf{p}$ is the particle momentum, $e$ is the particle charge, $\mathbf{E}$ is the $E$-field, $\mathbf{B}$ is the $B$-field, $m$ is the particle mass, and $c$ the speed of light in a vacuum. Using Equation~(\ref{Lorentz}) for the equations of motion assumes that the particle densities are low enough that the Lorentz force dominates over other forces, namely inertia and pressure. To obtain Equation~(\ref{Lorentz}), one uses ${\bf p}\equiv\gamma m \bf{v}$, with
\begin{equation} \label{Lor_fac}
\gamma = \frac{\sqrt{m^{2}c^{4} + {\bf p}^{2}c^2}}{mc^{2}},  
\end{equation}
as defined by the total particle energy $E_{\rm tot} = \gamma mc^{2}=\sqrt{m^{2}c^{4} + {\bf p}^{2}c^2}$. Using the momentum form of the Lorentz force equation makes it easier to include relativistic effects. To solve the ordinary differential equation (ODE) in Equation (\ref{Lorentz}), we compare various explicit higher-order schemes with embedded lower-order errors in the Runge-Kutta family of solvers as well as a symplectic scheme for comparison. For these explicit schemes, the order of the accuracy of the scheme is given, with the order of the error estimation given in brackets, and the stage indicating the number of function evaluations required. The numerical schemes that we compared are the Runge-Kutta Fehlberg 4(5)(5 stage) \citep{DORMAND80}, Verner's pair method DVERK 6(5)(8 stage) \citep{Hairer93}, Dormand-Prince 8(7)(12 stage) \citep{PRINCE81}, an adapted Curis 10(8)(18 stage) pair scheme, 
and an adapted Hiroshi 12(9)(29 stage) pair scheme. The Curtis 10(8) scheme is adapted from the original explicit Curtis $10^{\rm th}$-order scheme \citep{Hairer1978} to have an embedded $8^{\rm th}$-order error. Similarly, the explicit Hiroshi $12^{\rm th}$-order scheme \citep{Hiroshi2006} is adapted to have a $9^{\rm th}$-order error. The higher-order schemes' accuracy comes at the cost of more function evaluations; thus for a single time step, the higher-order schemes require many more computations. The advantage of the higher-order schemes, however, is that one can use larger time steps and achieve the same accuracy as the lower-order schemes, implying fewer iterations over the specified simulation scale, therefore saving computational time. The mentioned explicit schemes are also pair schemes, thus they have embedded lower-order schemes to easily calculate the truncation error needed to calculate the adaptive step size. Thus one does not have to rerun all the function evaluations for a halved/lower step size to obtain the truncation error, making them much more efficient. For a general implementation of the Runge-Kutta family of solvers and how these schemes are derived, see \citet{Shampine1977, Shampine1979}. 

A common problem is that ODE solvers become inaccurate and unreliable if the ODE is too stiff, namely if the function values change drastically. If small enough time steps are taken, this problem is avoided and the higher-order solvers seem to be less affected by the stiffness of the ODE due to their higher accuracy \citep{Ketcheson2013}.

Finally, we also implement the second-order Vay symplectic scheme \citep{Vay2008}, which conserves energy and is modified to work better for large field strengths than the more common Boris push symplectic scheme \citep{Boris1970}. The symplectic integrators are valuable as benchmarks, since they are implicit, stable, and are derived from the Hamiltonian to conserve phase space. A drawback of the symplectic schemes is that they require a constant time step or manually changing the time steps. The main problem with the use of symplectic integrators for our purposes is that they are derived by solving the Hamiltonian, with energy conservation, thus the RRF can not be included and still conserve energy. To add the RRF to a symplectic integrator, the Hamiltonian would have to be re-derived and solved from the start; one can not just subtract the RRF from the symplectic result. A few authors have used such a symplectic integrator with the RRF added from \citet{Tamburini2010} in their PIC calculations, but they did not re-derive the corresponding Hamiltonian. It has not been demonstrated that the scheme conserves phase space volume for the scheme to be considered truly symplectic. Without energy conservation, it would be similar to using a low-order scheme with the same order error as the symplectic integrator, namely second order for Vay. 

We abbreviate the schemes that we have implemented for the rest of the paper as follows: 
\begin{itemize}
    \item Runge-Kutta-Fehlberg \citep{DORMAND80}: RKF
    \item Verner's pair method \citep{Hairer93}: DV
    \item Dormand-Prince \citep{PRINCE81}: PD
    \item Curis pair scheme \citep{Hairer1978}: CR
    \item Hiroshi pair scheme \citep{Hiroshi2006}: HR
    \item Vay \citep{Vay2008}.
\end{itemize}

\subsection{Adaptive Time Step}\label{sec:2.1}
The advantage of using higher-order adaptive solvers is that they can maintain stability using larger time steps. This is crucial, since each time step, whether using a symplectic, implicit, or explicit scheme, introduces inaccuracy. Thus it is beneficial to minimise the number of time steps as much as possible while maintaining a reliable accuracy. Another benefit is saving computational time in a dipole-like or spatially changing $B$-field structure, where one can take larger time steps as the particle gyro-radius increases. For our implementation, we use the chosen explicit schemes that have an embedded lower-order result, which allows for an efficient and less computationally heavy calculation of the truncation error $\tau_{\rm{err;n}}$ that is used to calculate an adaptive time step. 

To calculate the truncation error, we use the higher-order result for the particle momentum $p_{\rm h}$ and the lower-order result $p_{\rm l}$ and normalise it:
\begin{equation}
\tau_{\rm err} = \frac{\vert p_{\rm h} - p_{\rm l} \vert}{p_{\rm h}}.
\end{equation}
Using $\tau_{\rm{err}}$, we can calculate an adaptive time step as follows \citep{SODERLIND2006}:
\begin{equation} \label{t_adap}
\Delta t_{n+1} = \left(\frac{\Delta t_{n}}{\Delta t_{n-1}}\right)^{-\frac{1}{b}} \left(\frac{TOL}{\tau_{\rm{err;n}}}\right)^{-\frac{1}{b\left(p+1 \right)}} \left(\frac{TOL}{\tau_{\rm{err;n-1}}}\right)^{-\frac{1}{b\left(p+1 \right)}},
\end{equation}
where $\Delta t$ is the time step, $TOL$ is the chosen tolerance for $\tau_{\rm{err}}$, $b \in \left[2,8 \right]$, $n$ indicates the $n^{\rm th}$ time step, and $p$ is the order of the chosen numerical method. Additionally, we tested various adaptive time-step calculations from \citet{Soderlind2003} to find a suitably accurate and stable method, since the standard method \citep{PRINCE81} introduced instabilities, especially when introducing $E$-fields. The standard method only uses the previous time step and truncation error whereas Equation~(\ref{t_adap}) uses the 2 prior time steps and truncation errors.

To constrain how much the new step can increase or decrease while smoothly transitioning to a stable result instead of overshooting, we introduce a limiting function. \citet{SODERLIND2006} provide an equation for limiting the new time step $\Delta t_{\rm l}$:
\begin{equation}
\Delta t_{\rm l} = \Delta t_{n}\left[1 + \kappa \arctan\left(\frac{\Delta t_{n+1} - \Delta t_{n}}{\kappa \Delta t_{n}} \right)  \right], 
\end{equation}
with $\kappa \in \left[0.7,2.0 \right]$.
We find that using $\kappa=0.7$ for all schemes, $b=8$ for RKF and DV, and $b=4$ for PD, CR, and HR yield the most stable results. On top of these steps, we add a selection filter by checking that each momentum component's truncation error is below the specified tolerance $TOL$, otherwise, we repeat this process with the newly obtained time step until the truncation error is sufficiently small. We also investigated the addition of an error estimator constant $\rho$ to these equations, similar to \citet{Noventa2018}, but found no major increase in stability, thus we omitted it for simplicity.

After the described steps, we obtain the updated momentum components with which we can calculate the new Lorentz factor using Equation~(\ref{Lor_fac}). We then calculate the particle's new position by solving $d\mathbf{x}/dt = \mathbf{v} = \mathbf{p}/\gamma m$ using the Euler method. The reason for using the Euler method is that the desired accuracy was already obtained with the first integration, and there was no noticeable difference when using a higher-order method again, except for an increase in computational time. This is similar to what was done by \citet{Cerutti2013, Petri2017, Ripperda2018}. 

Note that since the symplectic integrator requires a constant time step, we compare the results from the explicit schemes to those of the symplectic integrator using a constant time step for these schemes. When calculating the results for the dipole field scenarios, we change the time step by ensuring that we sufficiently sample the relativistic gyro period
\begin{equation}
P_{\rm g} = \frac{2\pi\gamma m}{\vert e \vert B}.
\end{equation}
This yielded stable results for all the methods and allowed us to change the time step for the symplectic scheme. The difficulty with a fixed-step scheme is knowing if $P_{\rm g}$ or the gyro-radius is sufficiently sampled. This changes for each scenario, parameter setup, and as the variables change over time. This is where the adaptive schemes have a major advantage in identifying a suitable time step. 

\subsection{Particle Set-up and Dipole $B$-field}\label{sec:2.2}
For this paper, we use a static vacuum dipole $B$-field for some of our test scenarios. To calculate the $B$-field, we use Equations~$3.11 - 3.13$ from \citet{Dyks2004} for a retarded dipole $B$-field. By setting the parameter $r/R_{\rm LC}=0$, these equations reduce to the static dipole $B$-field, where $R_{\rm LC}$ is the light cylinder radius. This radius, where the corotation speed equals that of light in vacuum, can be calculated using $R_{\rm LC}=cP_{s}/2\pi$, where $P_{\rm s}$ is the spin period of the star.\footnote{The more general retarded dipole equations were used since we aim to implement them in later work and they represent more realistic stellar $B$-fields than the static dipole case.} For the $E$-field inside the light cylinder, we use the perpendicular $E$-field from \citet{GJ1969},
\begin{equation}\label{E-Dipole}
\mathbf{E}_{\perp} = -\frac{\mathbf{\Omega} \times \mathbf{r}}{c} \times \mathbf{B},   
\end{equation}
where $\Omega$ is the angular velocity of the star and $r$ the radial position. 

It is important to note that linear interpolation of a numerical solution of electromagnetic fields on a grid does not necessarily preserve the non-divergent nature of the $B$-field \citep{Schlegel2020}. This effect is on a scale smaller than the resolution of the $B$-field and is negligible for high-resolution magnetic fields. This does not affect the analytical fields in this paper, but is a consideration for our next work where we will be using the force-free field grids similar to what was done by \cite{Harding2021}.   

To solve Equation~(\ref{Lorentz}), we need to specify the initial particle position and momentum. For the dipole $B$-field scenarios, we set up a simple procedure to specify the initial parameters but it is usable for any case. We specify a Lorentz factor and pitch angle\footnote{We define `pitch angle' $\theta_{\rm p}$ as the angle between the particle's velocity and the local $B$-field direction.} for the particles at the initial position. We start by defining a momentum vector in Cartesian coordinates, with the magnitude determined by the Lorentz factor. The momentum vector is tilted from the $z$-axis by the given pitch angle, with the azimuthal phase initially set to zero. Since the Cartesian $B$-field components can be calculated, we can calculate the rotation angles to align the momentum vector with the $B$-field direction. We define the following rotation matrix to rotate the momentum vector towards the magnetic field direction:
\begin{equation}
R_1 = 
\begin{bmatrix}
\cos\phi\cos\theta & -\sin\theta & \cos\phi\sin\theta \\
\sin\phi\cos\theta & \cos\theta & \sin\phi\sin\theta \\
-\sin\theta & 0 & \cos\theta
\end{bmatrix}.
\end{equation}   
In this matrix, $\theta$ is the pitch rotation and $\phi$ is the yaw rotation. This then sets up the particle with the specified Lorentz factor and a pitch angle ($\theta_{\rm p}$) relative to the local $B$-field.        

\subsection{$\mathbf{E}\times \mathbf{B}$-Drift Frame Calculations}
Upon introducing an $E$-field, one needs to calculate some quantities in the $\mathbf{E}\times \mathbf{B}$-drift velocity co-moving frame for the accuracy and stability tests. These values include the Lorentz factor and relativistic gyro-radius. Using the drift velocity for uniform fields $\mathbf{V}_{EB}=c\mathbf{E}\times\mathbf{B}/B^{2}$ and the corresponding drift Lorentz factor $\gamma_{\rm d}$, we perform a general Lorentz transformation given by the four-vector matrix \citep{Furry1955}:
\begin{equation}
\footnotesize
\boldsymbol{\stackrel{\leftrightarrow}{T}} \, = 
\begin{bmatrix}
\gamma_{\rm d} & -\gamma_{\rm d}V_{x} & -\gamma_{\rm d}V_{y} & -\gamma_{\rm d}V_{z} \\
-\gamma_{\rm d} V_{x} & 1+(\gamma_{\rm d} -1)\frac{V_{x}^{2}}{V^{2}} & (\gamma_{\rm d} -1)\frac{V_{x}V_{y}}{V^{2}} & (\gamma_{\rm d} -1)\frac{V_{x}V_{z}}{V^{2}} \\
-\gamma_{\rm d} V_{y} & (\gamma_{\rm d} -1)\frac{V_{x}V_{y}}{V^{2}} & 1+(\gamma_{\rm d} -1)\frac{V_{y}^{2}}{V^{2}} & (\gamma_{\rm d} -1)\frac{V_{y}V_{z}}{V^{2}} \\
-\gamma_{\rm d} V_{z} & (\gamma_{\rm d} -1)\frac{V_{x}V_{z}}{V^{2}} & (\gamma_{\rm d} -1)\frac{V_{y}V_{z}}{V^{2}} & 1+(\gamma_{\rm d} -1)\frac{V_{z}^{2}}{V^{2}} 
\end{bmatrix},    
\end{equation}
where $V_{x}$, $V_{y}$, and $V_{z}$ are the drift velocity components of $V=V_{EB}$ in this matrix and $\gamma_{\rm d}$ the drift velocity Lorentz factor. Thus with $\boldsymbol{\stackrel{\leftrightarrow}{T}}$ we can obtain the transformed velocity and position. 
To transform the fields, we use the following set of equations \citep{Chow2006}:
\begin{equation}
\begin{aligned}
    \mathbf{B}_{\parallel}' =~ & \mathbf{B}_{\parallel} \\
    \mathbf{E}_{\parallel}' =~ & \mathbf{E}_{\parallel} \\
    \mathbf{B}_{\perp}' =~ & \left(\mathbf{B}_{\perp} - \beta_{EB}\times \mathbf{E}\right) \\
    \mathbf{E}_{\perp}' =~ & \left(\mathbf{E}_{\perp} + \beta_{EB}\times \mathbf{B}\right),
\end{aligned}
\end{equation}
where the perpendicular and parallel components are with respect to $V_{EB}$, and $\bf{\beta}_{EB} = \mathbf{V}_{EB}/c$. With these transformed values we can calculate the transformed frame Lorentz factor using $\gamma' = \gamma_{\rm d}\gamma \left( 1 - \mathbf{V}_{EB}\cdot \mathbf{V}/c^{2}\right)$ and the relativistic gyro-radius with the standard equation
\begin{equation} \label{gyro_rad}
R_{\rm g} = \frac{\gamma v_{\perp} mc}{eB},    
\end{equation}
where the $v_{\perp}$ component is defined with respect to the local $B$-field. 

\subsection{Radiation-Reaction Force} \label{sec:2.4}
To ensure that we adequately take the effect of radiative losses on the particle momentum into account, we include the generic RRF into Equation (\ref{Lorentz}). We used the fully generalised classical form for the RRF from \citet{landau1975} written in terms of the particle momentum\footnote{We also implemented the super-relativistic RRF equations from \citet{landau1975} but found them to be unstable when resolving the particle gyrations, thus these results were excluded.}:
\begin{equation} \label{Landau}
\begin{aligned}
\mathbf{f} =& ~\frac{2e^{3}\gamma}{3mc^{3}}\left\lbrace\left(\frac{\partial}{\partial t} + \frac{\mathbf{p}}{\gamma m}\cdot\nabla \right)\mathbf{E} + \frac{\mathbf{p}}{\gamma mc}\times\left(\frac{\partial}{\partial t} + \frac{\mathbf{p}}{\gamma m}\cdot\nabla \right)\mathbf{B} \right\rbrace \\
+&\frac{2e^{4}}{3m^{2}c^{4}}\left\lbrace \mathbf{E}\times\mathbf{B} +\frac{1}{\gamma mc}\mathbf{B}\times\left(\mathbf{B}\times\mathbf{p} \right) +\frac{1}{\gamma mc}\mathbf{E}\left(\mathbf{p}\cdot\mathbf{E} \right) \right\rbrace \\
-&\frac{2e^{4}\gamma}{3m^{3}c^{5}}\mathbf{p}\left\lbrace \left(\mathbf{E} + \frac{\mathbf{p}}{\gamma mc}\times\mathbf{B} \right)^{2} - \frac{1}{\gamma^{2}m^{2}c^2}\left(\mathbf{E}\cdot\mathbf{p} \right)^2 \right\rbrace. 
\end{aligned}  
\end{equation}

This classical equation holds if the relativistic gyro-radius of the particle $R_{\rm g}$ is larger than the electron Compton length scale $(h/mc)$. In the particle rest frame, it is required that the RRF $\ll$ Lorentz force, but in the observer frame the RRF can be equal to or exceed the Lorentz force \citep{landau1975, Vranic2016}. A limit can also be set where quantum effects need to be accounted for, namely where the field experienced by the particle $E_{f} = \vert \mathbf{p}\times \mathbf{B} \vert /mc $ is below the Schwinger limit $E_S = mc^2 /\vert e\vert \lambdabar_c$, with $\lambdabar_c$ the Compton wavelength \citep{Sokolov2010}. For more detail on the Schwinger limit and a review of the physics of highly magnetised neutron stars, see \citet{Harding2006}. At this limit, quantum effects affect the synchro-curvature radiation as well \citep[see also,][]{2017PhRvD..95j5008V}.   

The first term in Equation~(\ref{Landau}) is the contribution from the temporal and spatial changes in the fields. Calculating the spatial derivatives of all the field components is computationally expensive. Thus, similar to \citet{Cerutti2016, Tamburini2010}, we evaluated the relative contribution of this first term and found it to be negligible. A more extensive study was done by \citet{Vranic2016} comparing different forms and normalisations of Equation~(\ref{Landau}). They found that even in extreme field cases neglecting the first term gives similar results; however, when dealing with the equation stochastically in the quantum limit, this term should be included. Thus unless your field is changing rapidly spatially or temporally, the first term can be safely neglected. We also evaluated the different forms of the different classical RRF equations from \citet{Vranic2016} and found the results to be similar\footnote{There is a slight computational time advantage when using Equation~(\ref{Landau}), since we are using momentum in Equation~(\ref{Lorentz}) as well, thereby avoiding repetitive conversion calculations.}.    

Given the RRF, we can calculate the power radiated by the particle \citep{Jackson1975}
\begin{equation} \label{P_rad}
P_{\rm{rad}} = \mathbf{F}_{\rm{rad}}\cdot\mathbf{v},
\end{equation} 
where $\mathbf{F}_{\rm{rad}}$ is the self-force experienced by the particle due to radiating, namely the RRF. The energy radiated can then be calculated using \citep{Jackson1975}
\begin{equation} \label{E_rad}
E_{\rm{rad}} = \int^{t_{1}}_{t_{2}}(\mathbf{F}_{\rm{rad}}\cdot\mathbf{v}) dt.
\end{equation}
Using Equation~(\ref{E_rad}), the total radiated energy can be calculated. To see if our numerical calculations are self-consistent, we add the particle energy and radiated energy to obtain the total energy; and we evaluate deviations from the initial total energy\footnote{This only holds if there is no $E_{\parallel}$-field to accelerate the particle. If there is an $E_{\perp}$-field but no $E_{\parallel}$-field parallel to the local $B$-field, the calculations can be done in the frame co-moving with the $\mathbf{E}\times\mathbf{B}$-drift velocity.}.    

\subsection{Comparison to Aristotelian Electrodynamics (AE)} \label{sec:2.5}
In the case of extreme radiation reaction, as is thought to be applicable in pulsar magnetospheres, the particles' dynamics are governed by the equilibrium of the Lorentz force and the RRF. Since the particles are accelerated to extreme Lorentz factors, the particles are assumed to be light-like and are assumed to follow the null geodesic. Such a limit defines AE, which is crucial in understanding pulsar particle dynamics for gamma-ray emission \citep{Kalapotharakos2019, Gruzinov2013}. \citet{Yangyang2022} summarises AE as well as highlighting challenges with some of the proposed AE models, since it has not yet concretely been confirmed.

AE seems to have been first proposed by \citet{Finkbeiner1989} and later popularised by \citet{Gruzinov2012}. We will continue to discuss AE in more detail in follow-up work, more specifically the application to non-uniform fields namely the retarded vacuum dipole- and force-free fields. AE assumes the particles are travelling along the principal null directions, where the leading term of the RRF determines the particle trajectory as it travels at the speed of light. The particle trajectory is described as \citep{Gruzinov2012}:
\begin{equation} \label{AE}
\mathbf{v}_{\pm} = \frac{\mathbf{E}\times \mathbf{B} \pm \left( B_{0}\mathbf{B} + E_{0}\mathbf{E} \right)}{B^{2} + E_{0}^{2}}c,
\end{equation}
where $\pm$ is for a positive or negative charge and $E_{0}$ and $B_{0}$ are invariants. These invariants\footnote{Since $Q$ and $P$ are scalar invariants of the electromagnetic field tensor, so are the scalars $E_0$ and $B_0$.} are calculated using
\begin{equation}
\begin{aligned}
E_{0} &= \sqrt{\sqrt{\left(P/2\right)^{2} + Q^2} - P/2} \\
B_{0} &= \frac{Q}{|Q|} \sqrt{\sqrt{\left(P/2\right)^{2} + Q^2} + P/2},
\end{aligned}    
\end{equation}
where $P=B^{2} - E^{2}$, $Q=\mathbf{E}\cdot\mathbf{B}$, and $E_{0}>0$. This argues that the velocity $c(\mathbf{E}\times\mathbf{B})/(B^{2}+E^{2}_{0})$ transforms to an arbitrary Lorentz frame where the $E$-field is parallel to the $B$-field, so that there is no $E_{\perp}$ component to drift the particle off the local $B$-field. The $E_{0}$ addition to the drift velocity means the drift velocity does not diverge beyond $R_{\rm LC}$ as the $B$-field decreases in a dipole-like $B$-field structure (i.e., as $r^{-3}$). Importantly, AE requires the particle to be rapidly accelerated to a critical Lorentz factor obtained by balancing the power due to the accelerating $E$-field with the energy lost due to curvature radiation \citep{Gruzinov2013},
\begin{equation}
ceE_{0} = \frac{2e^{2}c\gamma^{4}}{3R_{\rm c}}.    
\end{equation}
This yields
\begin{equation} \label{gam_crit}
\gamma_{\rm c} = \left( \frac{3E_{0}R_{\rm c}^{2}}{2\vert e\vert }\right)^{\frac{1}{4}},    
\end{equation}
where $R_{\rm c}$ is the radius of curvature of the particle trajectory. Equation~(\ref{gam_crit}) assumes curvature radiation is the dominant loss process. If one balances the gained power with the synchro-curvature radiated power, one would have a more general expression transitioning between the synchrotron and curvature regimes. Equation~(\ref{AE}) describes the particle's trajectory as the particle travels out to infinity, implying that inside the pulsar magnetosphere ($r<R_{\rm LC}$) a large accelerating $E$-field is required to accelerate the particle to the critical Lorentz factor within a short time scale for these equations to be applicable close to the pulsar surface. We define a deviation angle $\theta_{\rm D}$ as the angle between our model velocity and the AE limit velocity as defined in Equation~(\ref{AE}).

To calculate the radius of curvature, we use polynomial differentiation similar to what was done in \citet{Barnard2022}. We adopt this scheme, since in our next paper we will show our comparison calibration results with respect to their code. To calculate the second-order spatial derivative, we use
\begin{equation}
x''_{i} = \frac{\left(-3x'_{i-1} + 4x'_{i} - x'_{i+1} \right)}{2ds},
\end{equation}
where $x'$ is the first-order derivative and $ds$ is the step length (arclength) along the curved path. The radius of curvature can then be calculated using
\begin{equation}
R_{\rm c} = \frac{1}{\sqrt{(x'')^{2} + (y'')^{2} + (z'')^{2}}}.    
\end{equation}

\subsection{Calibration tests} \label{sec:2.6}
To test the accuracy of the various ODE solvers, we have set up standard scenarios with known analytical solutions. Apart from being able to compare with known solutions, these test cases are also applicable to a wide range of astrophysical problems and environments (e.g., local regions in a pulsar magnetosphere, or a magnetic mirror in a white dwarf magnetosphere).

(i)~The first scenarios using $\mathbf{E}=0$ (with and without an RRF) include:
\begin{itemize}
    \item A constant $B$-field.
    \item A static vacuum magnetic dipole.
\end{itemize}
In these listed scenarios, we compare the calculated relativistic gyro-radius to the analytical result from Equation~(\ref{gyro_rad}) and track any change in $\gamma$, since we expect it to stay constant for no RRF. Thus the relative error in $\gamma$ allows us to directly track the numerical error build-up in the results and the relative error in $\gamma$ per step allows us to assess the numerical error per step. For the static dipole case, we also consider the first adiabatic invariant $\mu =p_{\perp}^{2}/B$ to evaluate the accuracy of our magnetic mirroring result. For the RRF cases with $\mathbf{E}=0$, we use the same scenarios as without the RRF. However, we evaluate if our total energy is conserved for our accuracy tests, since $\gamma$ changes if RRF is included, and we compare our results to \citet{Petri2022} where possible. In the co-moving frame, we can once again compare to the analytical expression for $R_{\rm g}$, as well as monitor any change in $\gamma'$ in this frame. Some of these tests were similarly performed by \citet{Petri2017, Ripperda2018} and will serve as a comparison. 

(ii)~The test scenarios when including an $E$-field (constant $B$-field with a constant $E_{\perp}$-field; we set $E_{\perp}$ to a fraction of the $B$-field strength) are:
\begin{itemize}
    \item Zero initial momentum.
    \item A large initial Lorentz factor.
\end{itemize}
We also include the RRF, but with no analytical results available to serve as a comparison in this case.

(iii)~As another test and application to AR Sco, we set up the magnetic mirror scenario proposed by \citet{Takata2017} and compare our results to theirs. They use a static dipole $B$-field with no $E$-field, and a magnetic inclination angle of $\alpha=0^{\circ}$. They injected the particle at the companion centre with an initial $\gamma = 50$ and a pitch angle of $\sin\theta_{p\rm } = 0.1$. As far as we can tell, they constrain the particle to move along the $B$-field and simply push it along this field. The surface $B$-field is set using the magnetic moment as $\mu_{\rm WD}= 6.5\times 10^{34} \, \rm{G}\, \rm{cm}^{3}$, the WD radius as $R_{\rm WD} = 0.8R_{\odot}$, a binary separation of $a=8\times10^{10} \, \rm{cm}$, a WD spin period of $P_{\rm s} = 117$~s, and setting the orbital phase at $0.25$ where orbital phase $0.0$ would be at inferior conjunction. They adopt the transport equations from \citet{Harding2005}, but written in terms of the first adiabatic invariant in the observer frame, 
 \begin{equation} \label{Takata_transport}
 \begin{aligned}
 \frac{d\gamma}{dt} =& -\frac{P_{\perp}^{2}}{t_{\rm s}} \\
 \frac{d}{dt}\left(\frac{P_{\perp}^{2}}{B} \right) =& -2\frac{B}{t_{\rm s}\gamma}\left(\frac{P_{\perp}^{2}}{B} \right)^{2}  ,
 \end{aligned}
 \end{equation}
where $t_{\rm s} = 3m_{\rm e}^{3}c^{5}/2e^{4}B^{2}$ and $P_{\perp}=\gamma\beta\sin\theta_{\rm p}$. 
These equations assume that the particle is super-relativistic and has a small pitch angle. They set $E_{\parallel}=0$ since the $E$-field is assumed to be screened, but do not include an $E_{\perp}$. They thus neglect any $\mathbf{E}\times \mathbf{B}$-drift in the particle dynamics as the particle is just pushed along the $B$-field. Additionally, we reproduce their scenario but include an $E_{\perp}$-component to investigate the effect this would have on the results. 

(iv)~As a test for the RRF, we investigated the radiation-reaction limit and compared our results to the AE results in this regime. We compare our results to the limiting velocity in Equation~(\ref{AE}) as well as calculate the angle between our velocity vector and that of the AE velocity, as a deviation angle. For the AE results comparison we only investigate uniform fields in this paper, but with both $E_{\perp}$ and $E_{\parallel}$ components. In a follow-up work, we will investigate AE in retarded dipole- and force-free fields.

To find the best scheme for modelling AR Sco and pulsars, we balance sufficient accuracy (not introducing significant numerical errors in our calculations) with computational efficiency. We track the relative error in $\gamma$, the error in $\gamma$ with each step, and the relative error in total energy to investigate the numerical error build-up in the calculations. The relative error build-up should thus be low $(\ll 1)$ and the error introduced per step should be small enough that the amount of steps required for the simulation does not cause the errors to build up that would affect the results. We believe a relative error below $10^{-3}$ is ideal, but $10^{-2}$ is passable with some caution. An example relative error of $10^{-6}$ per step means after $\sim 10^{6}$ iterations the first significant figure is affected. Thus, one would be able to perform $10^{3}$ iterations in this case before hitting the $10^{-3}$ accuracy level. It is crucial for the numerical errors not to be exceedingly large since they could cause the particles' velocities to increase exponentially or have random fluctuations or oscillations. This would affect all the emission calculations and emission directions in the emission maps and spectra calculations making it impossible to differentiate between actual results and numerical errors. 

To evaluate the computational cost of each method while also evaluating their accuracy, we use the same amount of steps/iterations for each scheme for the uniform field cases\footnote{We use a set length scale for the dipole field cases.}. This means that if the total time simulated for one scheme is larger than another (or more gyrations occur for gyro-period normalised time) we know the latter scheme is taking larger time steps and will perform fewer iterations if given a set simulation time or length scale, which indicates a faster runtime for that scheme. We also track the time steps each scheme uses, directly showing which scheme is faster by using larger time steps combined with having the larger simulation time with the same amount of iterations. We also directly compare the schemes against one another in benchmark tables in section \ref{sec:3.3}. 

\section{Results}\label{sec:3}
In this section, we show the results for high field values and high particle energies, since these values are closer to standard pulsar parameters. This allows us to demonstrate that we have obtained sufficient accuracy and stability when using these extreme parameters. To assess stability, we investigate if the particle follows the expected trajectory (namely a deviation in $R_{\rm g}$) or if there is any numerical drift in expected values. The results are found to be more accurate and stable for lower field values, lower Lorentz factors, and small RRF scenarios. We run each method for the same number of iterations in each scenario, allowing us to assess the runtime of each method in each scenario. This also gives a good standard to indicate which schemes would be faster, given a set simulation time or length scale. While we used electrons for our simulations, any particle can be simulated for each scenario. 

\subsection{$B$-field scenarios with no RRF}\label{sec:3.1}
For the two $B$-field scenarios (constant and static vacuum dipole fields) we plot similar variables, thus to reduce repetition we clarify the details behind the plots first. For the trajectory plots, we show the 2D normalised position components. In the uniform field case, we plot the $xy$-plane and normalise to the $R_{\rm g}$ and in the dipole case we use the $xz$-plane, with position coordinates normalised to $R_{\rm{LC}}$. For the $\gamma_{\rm err}$, we plot the relative error in Lorentz factor $\gamma_{\rm err}=(\gamma - \gamma_{0})/\gamma_{0}$ vs time normalised to $P_{\rm g}$, allowing us to directly track the numerical error build-up. For the $R_{\rm g;err}$, we plot the relative error in gyro-radius $R_{\rm g;err} = (R_{\rm g} - R_{\rm g; an})/R_{\rm g; an}$ over normalised time. Here $\gamma_{0}$ is the initial Lorentz factor and $R_{\rm g; an}$ is the analytical gyro-radius. Finally, for the time step plots, we show the time steps taken over normalised time. In the static dipole field and RRF scenario, we do not normalise the time, since the gyro-period changes as the particle loses energy or the field strength changes. 

\begin{figure}
\centering
\includegraphics[width=.5\textwidth]{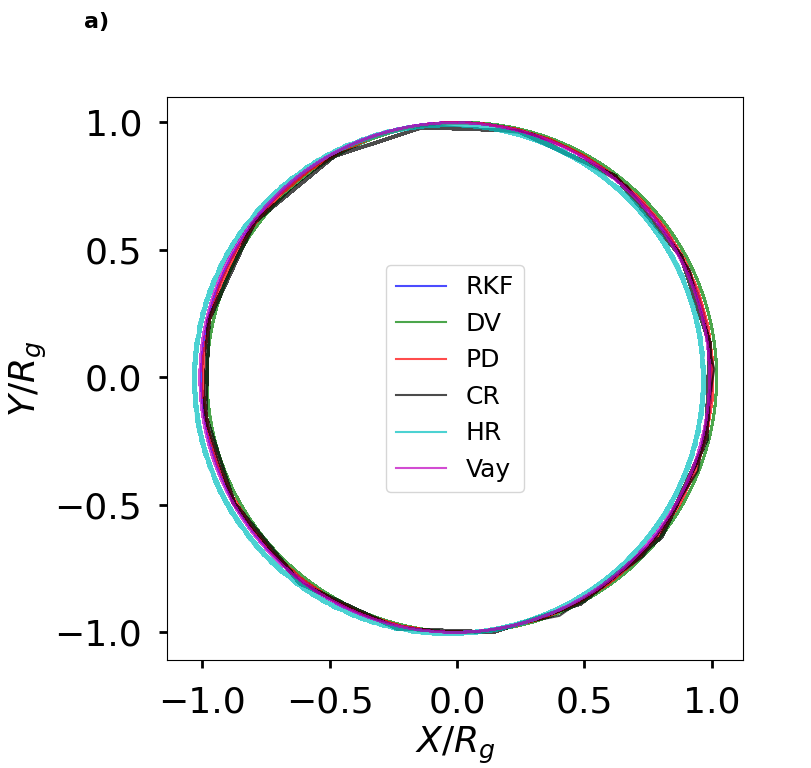}
\caption{The results for a constant and uniform $B$-field showing the particle trajectory in the $xy$-plane for each scheme. We use a constant time step of $dt = 10^{-13} \, \rm{s}$ for the Vay scheme.}
\label{Constant_B_rg}
\end{figure}

\begin{figure*}
\centering
\includegraphics[width=\textwidth]{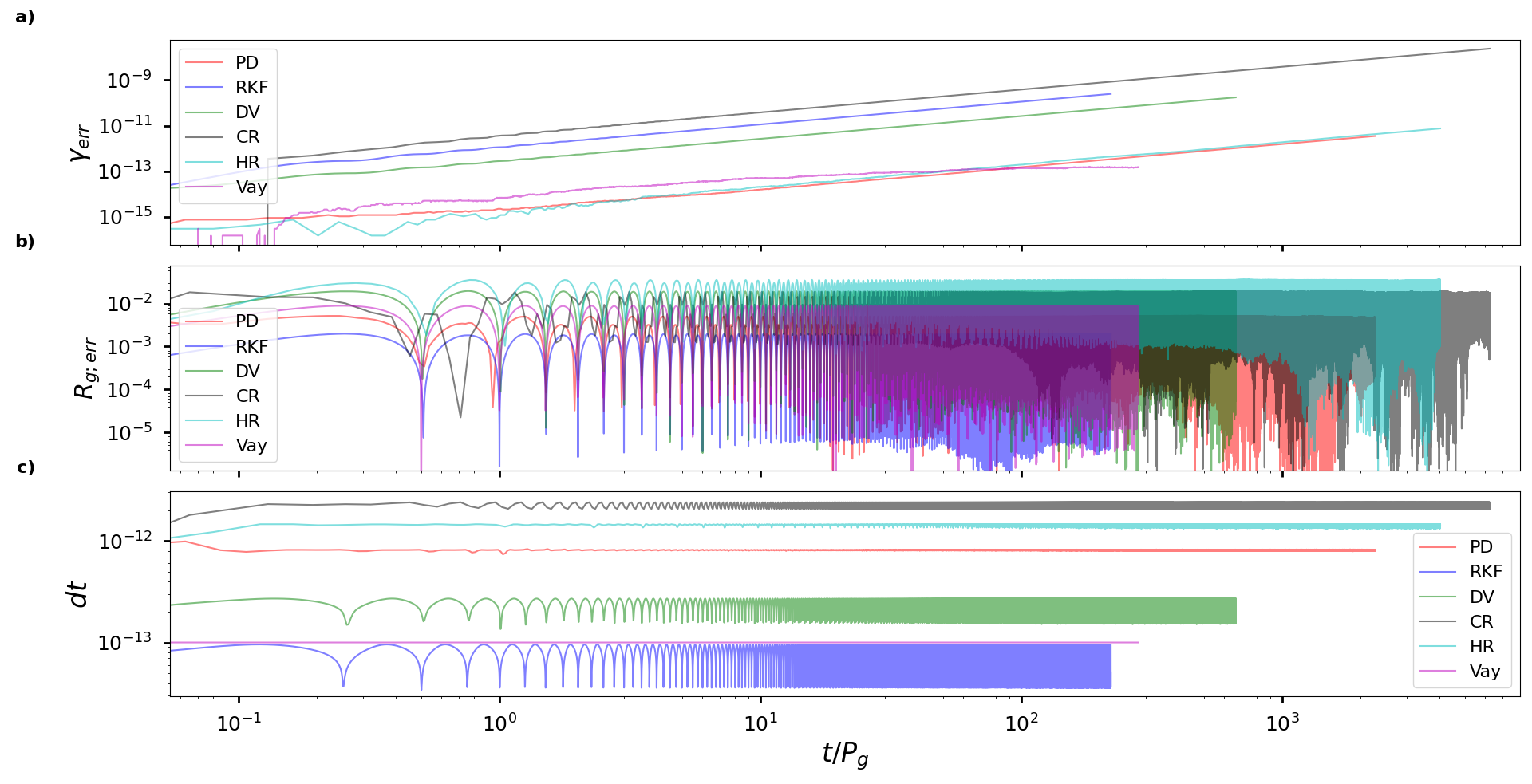}
\caption{The results for a uniform $B$-field for each scheme, where panel a) shows the relative error in $\gamma$, panel b) shows the relative error in $R_{\rm g}$, and panel c) shows the time step used. We use a constant time step of $dt = 10^{-13} \, \rm{s}$ for the Vay scheme.}
\label{Constant_B_err}
\end{figure*}

For the uniform $B$-field scenario, we used a $B$-field in the $z$-direction, namely $B_z = 10^{12} \,\rm{G}$. We set up our particle to start at the gyro-radius to gyrate around the point (0,0,0), with initial parameters $\gamma_{0} = 10^{8}$ and $\theta_{\rm p}=90^{\circ}$. In Figure~\ref{Constant_B_rg} and~\ref{Constant_B_err}, we compare the accuracy and stability of the various numerical schemes using the adaptive time steps as well as the Vay scheme using a constant time step of $dt = 10^{-13} \, \rm{s}$. 

In Figure~\ref{Constant_B_rg} we see that the schemes have some discrepancies in their resulting trajectories, where the CR and HR schemes have slightly shifted gyro-centres, but all schemes are stable over many gyrations. The roughness seen in the higher-order schemes' gyrations is due to time resolution, not inaccuracy in the schemes. 

In Figure~\ref{Constant_B_err}, panel a) we see that all the schemes have exceptionally small $\gamma_{\rm err}$ over multiple gyrations, with the cumulative error build-up visible as the relative error increases over time in the plot. The lower-order schemes have relative errors of $\gamma_{\rm err}\sim 10^{-9}$, the higher-order schemes $\gamma_{\rm err}\sim 10^{-11}$, and for the Vay scheme $\gamma_{\rm err}\sim 10^{-13}$. This indicates that the numerical errors are small enough to not affect the results within expected simulation runtimes of $10^{9}$ iterations as a high upper limit. It is evident that the CR scheme experienced a numerical instability, causing a large jump in the relative error. 

In panel b) of Figure~\ref{Constant_B_err} we see that the errors in $R_{\rm g}$ are relatively low for all schemes, but could potentially be a little higher than expected due to possible imperfect centring around (0,0,0). All the schemes except CR and HR have relative errors of $R_{\rm g; err}\sim 10^{-3}$, while the CR and HR schemes have $R_{\rm g; err}\sim 10^{-2}$. The results are found to oscillate to higher and lower values due to the adaptive time steps and possible minor instability in the gyro-radius as mentioned with the centring. When comparing the adaptive schemes to the Vay scheme, which is stable and reasonably accurate due to it being a symplectic scheme, the other schemes have similar error levels, giving confidence in the results. Similar results were also found by \citet{Ripperda2018} with somewhat lower relative errors in $R_{\rm g}$ at $R_{\rm g; err}\sim 10^{-5}$. Their error in $R_{\rm g}$ could be lower due to better centring or a better method of comparing the $R_{\rm g}$, but our errors in $\gamma$ are similar. Our results are stable over multiple gyrations, as noted from the trend of the errors, which seems to be quite flat. 

In panel c) of Figure~\ref{Constant_B_err}, we see a major advantage of the adaptive time step method, since it allows the higher-order schemes to take much larger time steps while maintaining the same accuracy as the lower-order schemes. These results show that the PD, CR, and HR schemes take $\sim 10$ times larger time steps than the RKF, implying that they complete 10 times more gyrations than the RKF with higher accuracy in the same number of iterations. It must be noted that the RKF does seem to be more stable than the higher-order schemes, which will be more apparent in the $E$-field results. This is most likely due to it requiring much smaller time steps than the other schemes.    

\begin{figure*}
\centering
\includegraphics[width=\textwidth]{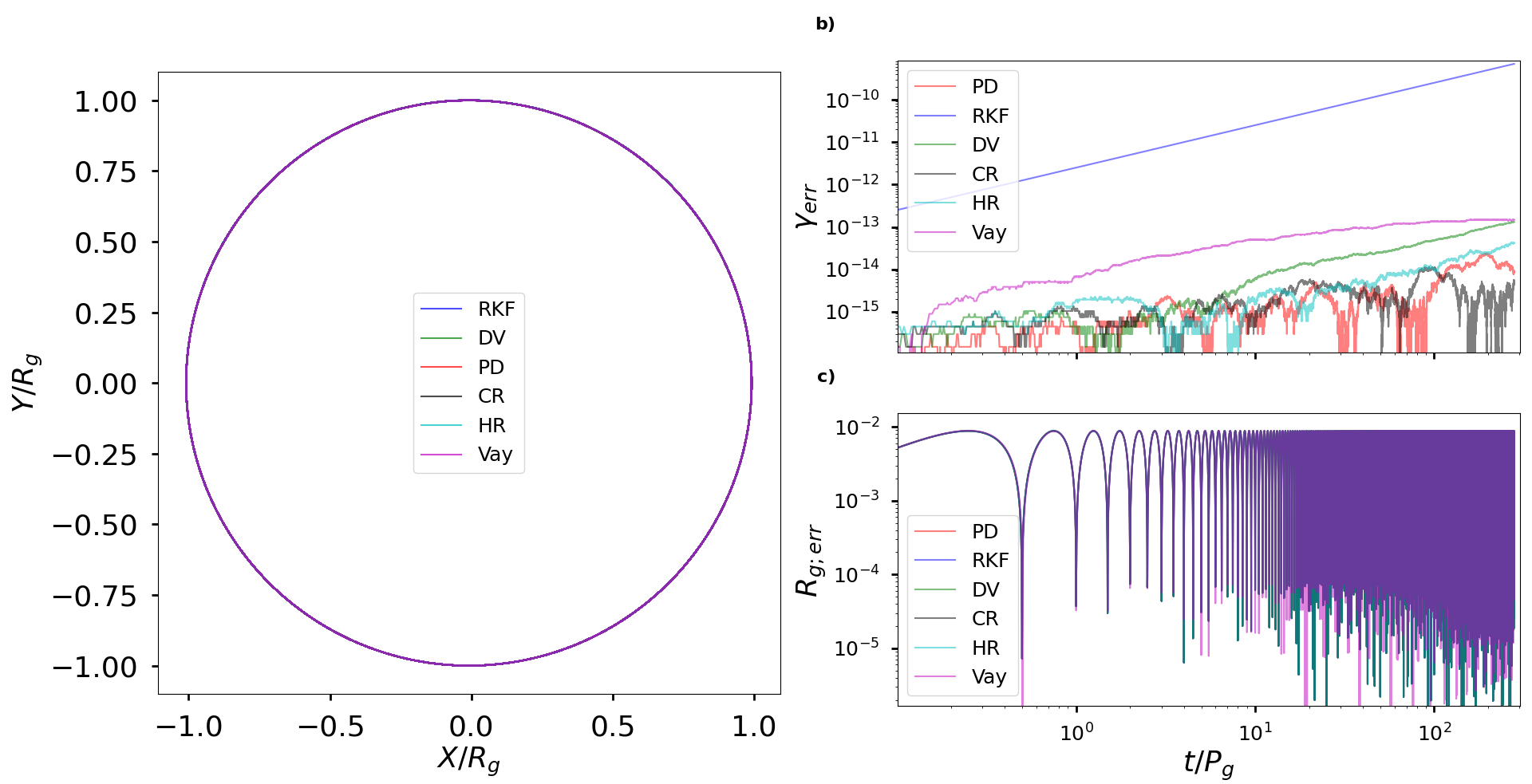}
\caption{Results for a constant $B$-field scenario as in Figures~\ref{Constant_B_rg} and~\ref{Constant_B_err}, but using a constant time step $dt = 10^{-13} \, \rm{s}$ for each method.}
\label{Constant_B_constant_dt}
\end{figure*}

In Figure~\ref{Constant_B_constant_dt} we consider the same constant $B$-field scenario, but now using a constant time step of $dt = 10^{-13} \, \rm{s}$ for each scheme. Panel a) indicates that all the schemes' trajectories are now much more stable and overlap one another on the plot. Panel b) shows that all the relative errors in $\gamma$ fall below the Vay scheme error of $\gamma_{\rm err}\sim10^{-13}$, except for the case of the RKF scheme. The RKF error is comparatively high, because $dt = 10^{-13} \, \rm{s}$ is fractionally above the stable time step for the RKF as seen in Figure~\ref{Constant_B_err} panel~c), so that the RKF would have significantly higher errors using this time step. In panel c) we see that all the schemes have similar relative errors in $R_{\rm g}$ just below $R_{\rm g; err}\sim 10^{-2}$, with quite stable results. 

Comparing the adaptive time step results in Figure \ref{Constant_B_err} with the constant time step results in Figure~\ref{Constant_B_constant_dt}, we see that decreasing the time step increases the accuracy of the schemes, as expected. The problem lies in identifying an acceptable time step before the run for an arbitrary scenario and setup parameters. This means that one would have to run each scenario multiple times to determine a suitable time step. It is then advantageous when an adaptive time step method can calculate a stable and accurate time step to use generally. The results of Figure~\ref{Constant_B_rg} and \ref{Constant_B_err} are without limiting the maximum time step of the adaptive schemes. Thus the results can be further improved by knowing the maximum stable time step for each scheme and for each scenario. This would once again require multiple runs to determine such a step, and we therefore only show the adaptive results, without setting a maximum to show we still obtain stable and accurate results.

\begin{figure*}
\centering
\includegraphics[width=\textwidth]{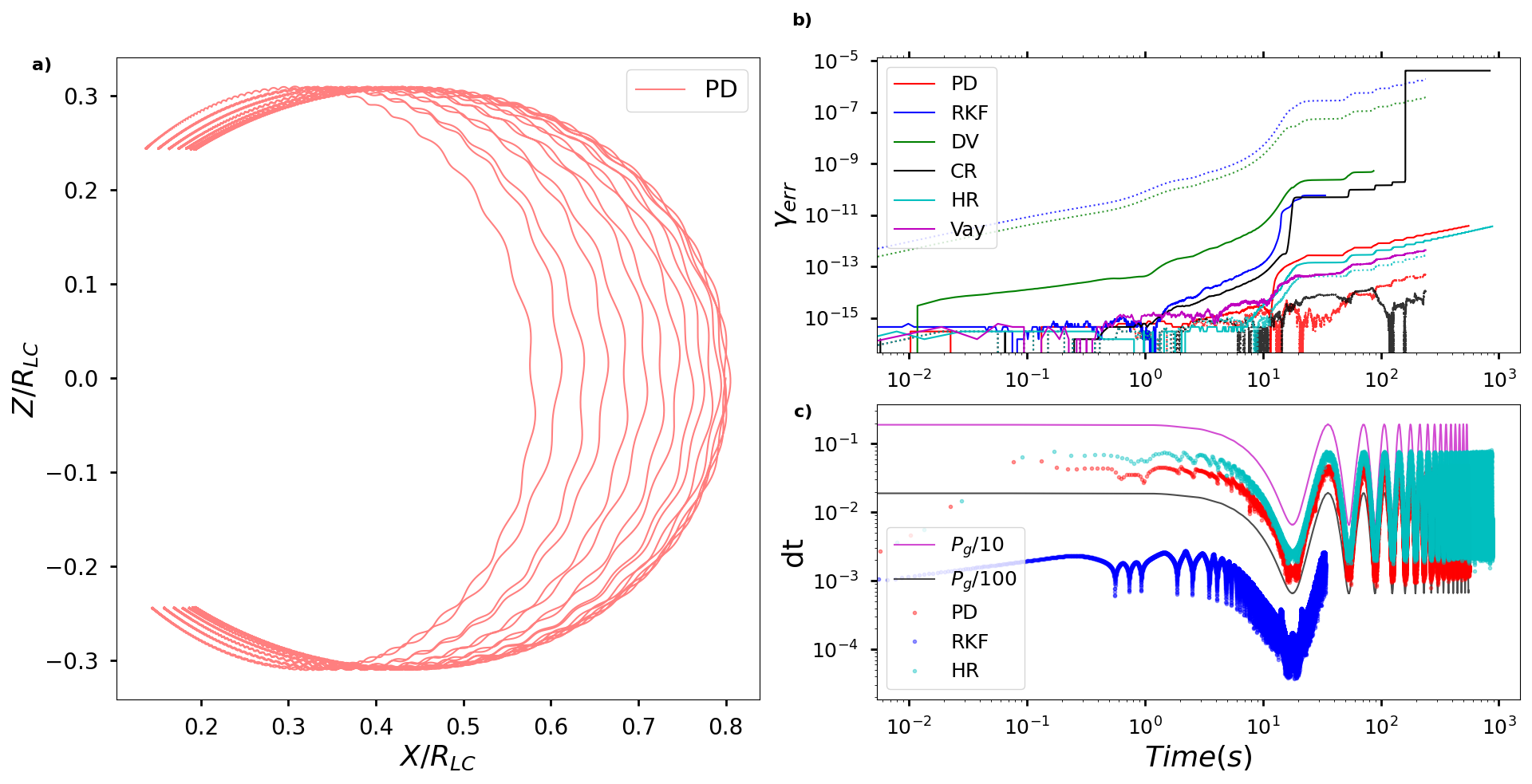}
\caption{Results for a static dipole $B$-field scenario where panel a) indicates the trajectory of the particle in the $xz$-plane. Panel b) shows the relative error in $\gamma$ where the dotted lines indicate when using time steps $dt = P_{\rm g}/100$. Panel c) shows the adaptive time step sampling for three of the schemes. We use the time steps $dt =  P_{\rm g}/100$ for the Vay scheme.}
\label{Dipole_B_large}
\end{figure*}

For the dipole $B$-field scenario, we use a surface $B$-field of $B_{\rm s} = 10^{10} \, \rm{G}$, a stellar radius $R_{\rm s} = 0.01R_{\odot}$, spin period $P_{\rm s}=117 \, \rm{s}$, and a magnetic inclination angle of $\alpha=0.0^{\circ}$. We initialise the particle at $0.8R_{\rm{LC}}$ with a Lorentz factor $\gamma_{0} =10^{8}$ and pitch angle $\theta_{\rm p} = 10^{\circ}$. In Figure~\ref{Dipole_B_large} panel a) we plot the trajectory of the particle in the $xz$-plane for the PD scheme as an illustration of the trajectory. We see the particle bounces between the magnetic mirrors at each magnetic pole and drifts in the $y$-direction due to the gradient drift, as expected. In panel b) we plot the $\gamma_{\rm err}$ for each scheme, where the dotted lines indicate using time steps of $dt = P_{\rm g}/100$ for each scheme, including Vay. The relative errors are found to be low for multiple magnetic mirrors for all the schemes, where the biggest jump in error happens at each mirror point. The adaptive PD and HR scheme's relative errors of $\gamma_{\rm err}\sim 10^{-11}$ are quite close in accuracy to those of Vay $(\gamma_{\rm err}\sim 10^{-12})$, but they are found to complete significantly more mirrors due to their larger time steps. The RKF and DV schemes' relative errors were found to be a little higher, but still accurate at $\gamma_{\rm err}\sim 10^{-10}$ and $\gamma_{\rm err}\sim 10^{-9}$, respectively. The panel further shows that the PD, CR, and HR schemes are more accurate than Vay when using the same time steps as illustrated by the dotted lines in the plot. The dotted RKF and DV lines yield higher relative errors because the time step used is higher than the stable time step required for these schemes. The adaptive CR is also seen to have experienced instabilities as indicated by the large jumps in its relative error. In panel~c) we can see how the adaptive time step changes, since the $B$-field changes as the particle bounces between the two poles and moves to different parts of the magnetosphere. The observed variation in the time steps is a factor $\sim 100$ for this case, implying that one can take a 100 times more large steps far away from the surface than when using a constant small time step closer to the surface, saving computation time. If one further probes with the adaptive time steps close to the stellar surface and far away, this variation can easily become a factor $10^6 - 10^{10}$. This is especially true for the pulsar case, where a particle at the surface moves outward beyond $R_{\rm{LC}}$. This highlights the crucial benefits of using an adaptive method in field configurations where one would expect more complicated field structures and variations. In panel~c) we also see that the RKF time steps are below $P_{\rm g}/100$, explaining why the dotted RKF and DV errors are so high in panel b). 

\begin{figure}
\centering
\includegraphics[width=.5\textwidth]{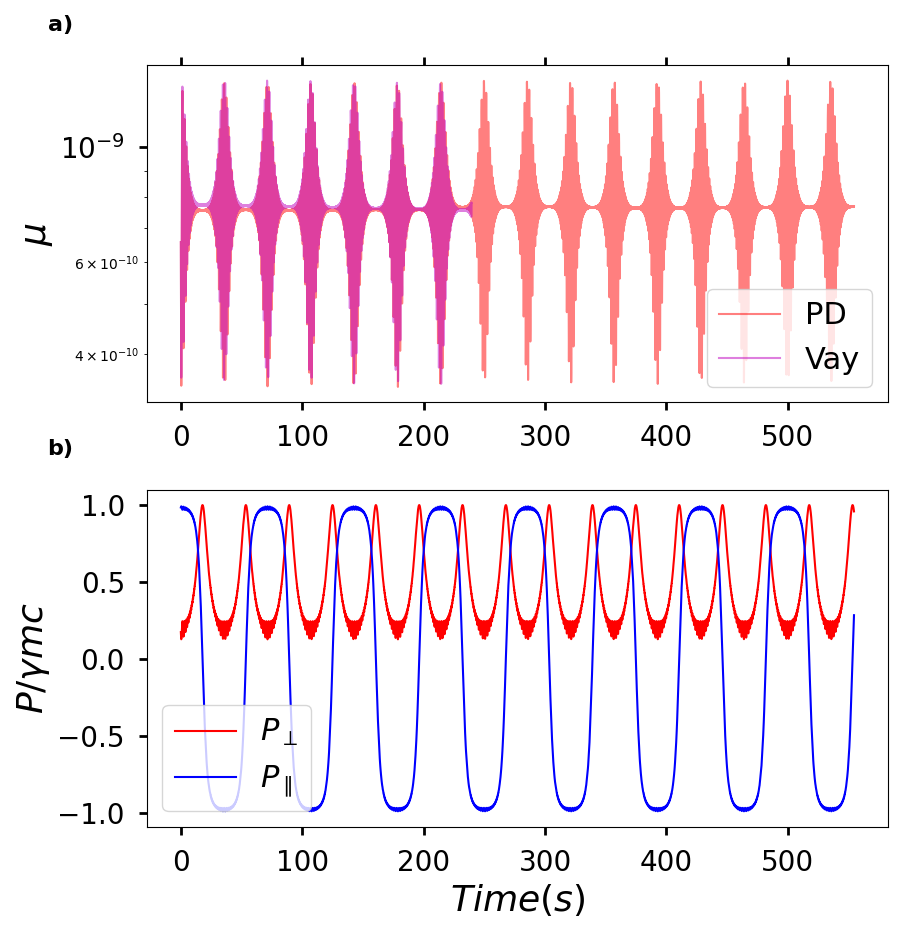}
\caption{Static dipole $B$-field scenario with panel a) showing the adiabatic invariant $\mu$ using $dt=P_{\rm g}/100$ for the Vay scheme. Panel b) shows $p_{\perp}$ in red and $p_{\parallel}$ in blue, both normalised by $\gamma mc$.}
\label{Dipole_B_large_mu}
\end{figure}

To illustrate that we are obtaining magnetic mirroring while maintaining accuracy, we plotted the first adiabatic invariant and momentum components. In Figure~\ref{Dipole_B_large_mu} we only show the results of the PD and Vay schemes in the first panel so as not to clutter the plot and only the PD in the second plot to show the momentum components. We find similar results to \citet{Soni2020} who simulated electrons in Earth's Van Allen belts, where one expects the average adiabatic invariant to stay constant. Thus even though the adiabatic invariant oscillates the average stays constant over many particle mirrors. This is found for both Vay and PD in our results. There is a minor offset seen between the Vay and PD schemes, which could be due to minor numerical instabilities. In panel b) we show $p_{\perp}$ and $p_{\parallel}$ normalised by $\gamma mc$ to identify the mirror points. The mirror points are located where $p_{\parallel}$ changes sign and where $p_{\perp}$ is a maximum, since this is where $\theta_{\rm p}=90^{\circ}$. Thus in this plot, one can see the particle bouncing between the magnetic poles multiple times.


\subsection{$B$-field RRF scenarios} \label{sec:3.2}
When calculating the relative error in energy $(E_{\rm err} = (E_{n} - E_{0})/E_{0})$ over time, we use as total energy $E_{n}$ the sum of the particle energy and the energy radiated by the particle and compare this to the particle's initial energy $E_0$. This allows us to self-consistently check if energy is conserved and if our results are sufficiently accurate.

\begin{figure}
\centering
\includegraphics[width=.5\textwidth]{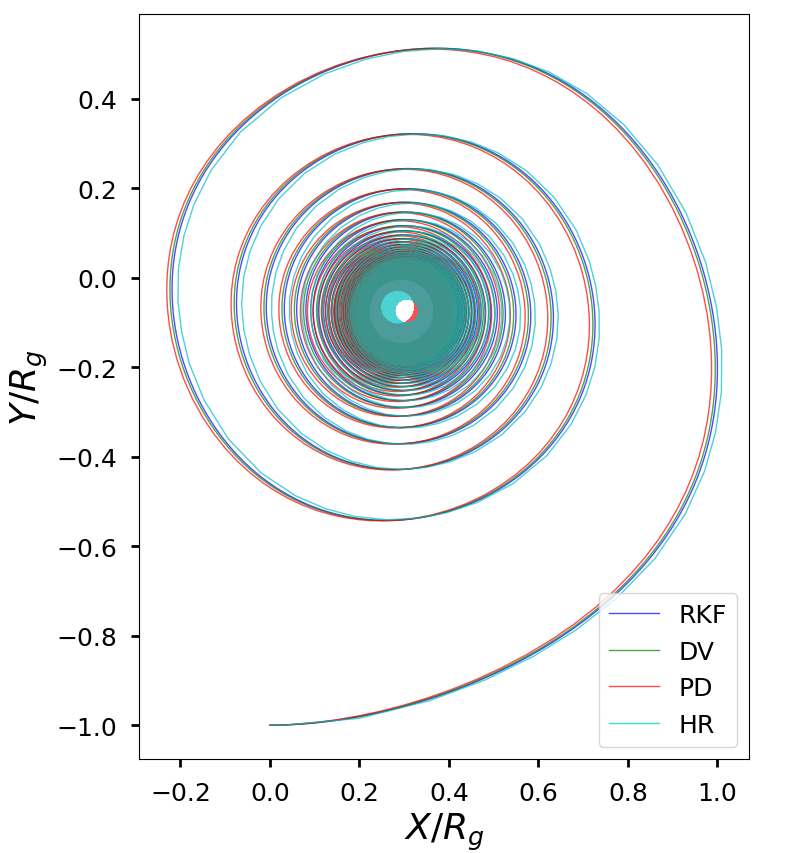}
\caption{Results for a constant $B$-field with the RRF included, showing the particle trajectory in the $xy$-plane for each scheme.}
\label{RRF_Constant_fields_rg}
\end{figure}

\begin{figure*}
\centering
\includegraphics[width=\textwidth]{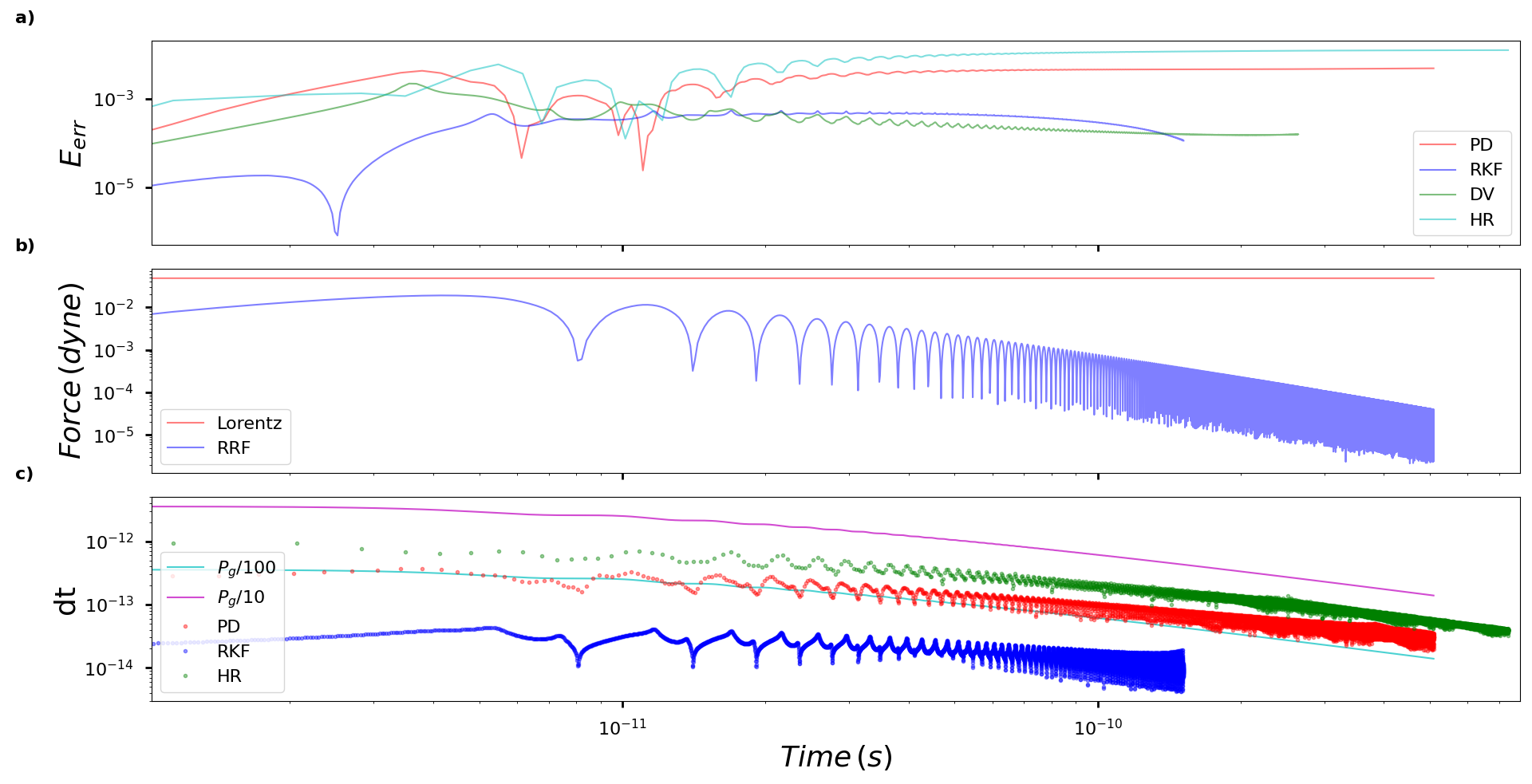}
\caption{Results for a constant $B$-field with the RRF included, where panel a) shows the relative error in total energy. Panel b) shows the Lorentz force in red and RRF in blue. Panel c) shows the adaptive time step sampling for three of the schemes where $P_{\rm g}/10$ is plotted in magenta and $P_{\rm g}/100$ in cyan.}
\label{RRF_Constant_fields_err}
\end{figure*}

For the uniform $B$-field RRF scenario, we use a $B$-field $B_{z} = 10^{8} \, \rm{G}$ with initial Lorentz factor $\gamma_{0} = 10^{4}$, pitch angle $\theta_{\rm p} = 90^{\circ}$ and included RRF. We used a scenario with a large radiation reaction, since these scenarios are the most unstable and inaccurate. In Figure~\ref{RRF_Constant_fields_rg} we see the particle gyrating around the $B$-field in the $xy$-plane, spiralling inwards as the particle loses energy due to the RRF, thus causing the gyro-radius to decrease. One can see some offset in the higher-order schemes but all the schemes show good stability except for CR (not indicated), which becomes unstable if we purely use the adaptive methods with no limits in time steps. In Figure~\ref{RRF_Constant_fields_err} panel a) we see that the adaptive schemes are accurate over many gyrations, with the RKF and DV schemes having higher accuracy $(E_{\rm err}\sim 10^{-4})$, since they are more stable even though they have lower-order accuracy. The PD has a relative error of $E_{\rm err}\sim 10^{-3}$ and the HR $E_{\rm err}\sim 10^{-2}$, with all the schemes producing accurate results over an extended simulation with multiple mirrors. Panel b) shows that for these parameters, the RRF is initially $\sim 10\%$ of the Lorentz force, but this decreases as the particle loses energy and the $\gamma$ decreases. This is quite high and is approaching the radiation-reaction limit regime, which we will be discussing in Section~\ref{sec:3.7}. This relatively high RRF causes the problem to become more stiff (i.e., rapidly changing variables), meaning that smaller time steps or higher-order schemes are required for adequate stability and accuracy. In panel c) we see that the higher-order schemes are able to use $\sim 10$ times larger time steps than the RKF, demonstrating that one can reach a much longer simulation time when using the same number of iterations.     

\begin{figure}
\centering
\includegraphics[width=.5\textwidth]{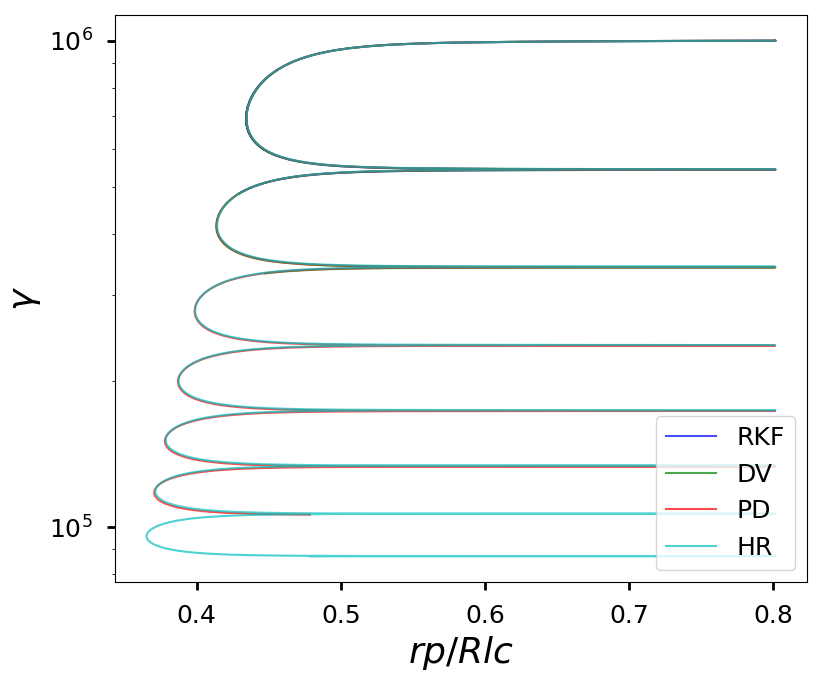}
\caption{Results for a static dipole $B$-field with RRF included, showing the Lorentz factor vs radius for each scheme.}
\label{RRF_Dipole_B_gam}
\end{figure}

For the static dipole scenario with RRF included, we used a surface $B$-field of $B_{\rm s} = 10^{8} \, \rm{G}$, stellar radius $R_{\rm s} = 0.01R_{\odot}$, spin period $P_{\rm s}=117 \, \rm{s}$, and a magnetic inclination angle of $\alpha=0.0^{\circ}$. We set the initial position of the particle at $0.8R_{\rm{LC}}$ with a Lorentz factor $\gamma_{0} =10^{6}$ and pitch angle $\theta_{\rm p} = 20^{\circ}$. In Figure~\ref{RRF_Dipole_B_gam} we plot the Lorentz factor vs the particle's radial distance normalised to $R_{\rm{LC}}$ for each scheme. One can see the Lorentz factor is slowly decreasing as the particle radiates energy until it is close to the mirror point, where it rapidly decreases as the particle radiates most of its energy in this relatively strong $B$-field. This therefore forms a step-like pattern as the particle is bounced between the stellar poles, since most of the energy is radiated at the mirror points due to the high local $B$-fields there. 

\begin{figure*}
\centering
\includegraphics[width=\textwidth]{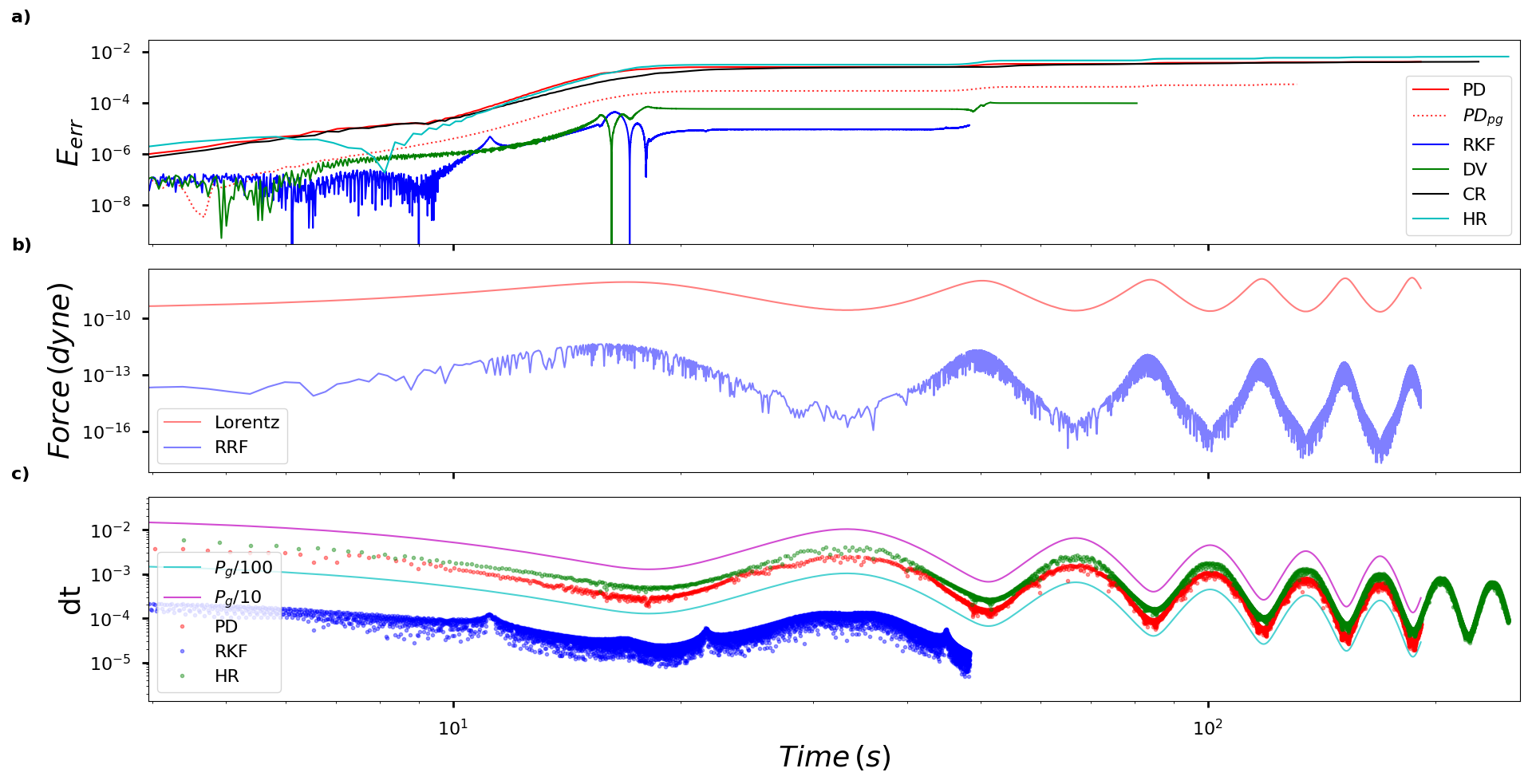}
\caption{Results for a static dipole $B$-field with RRF included. Panel a) shows the relative error in the total energy, with the dotted line indicating the use of $dt = P_{\rm g}/100$. Panel b) shows the Lorentz force in red and the RRF in blue, while panel c) shows the adaptive time step sampling for 3 of the methods.}
\label{RRF_Dipole_B_err}
\end{figure*}

In Figure~\ref{RRF_Dipole_B_err} panel a) we see quite small relative energy errors for each scheme, but the RKF and DV schemes were found to have the smallest errors due to them being more stable as mentioned in the previous scenario. The higher-order schemes have relative energy errors of $E_{\rm err}\sim 10^{-3}$ where the RKF has an $E_{\rm err}\sim 10^{-5}$ and the DV an $E_{\rm err}\sim 10^{-4}$. All the adaptive schemes thus have adequately small relative energy errors for this scenario. In this panel, we include a dotted line for the PD scheme using a time step $dt = P_{\rm g}/100$ to show that the schemes can be made significantly more accurate upon using smaller time steps or limiting the time steps, but one needs to know beforehand how small to make the time step. Using this time step, the relative error for the PD scheme was $E_{\rm err}\sim 10^{-4}$. In panel b) we show that in this scenario, the RRF$<0.01\%$ of the Lorentz force, indicating the methods' accuracy and stability in these high-RRF regimes. We also see increases in the RRF due to the higher $B$-fields at the poles and thus we expect most of the energy to be radiated at the mirror points. Notably, the small oscillations seen in the RRF curve are due to the gyration of the particle. Panel c) shows how the time steps become smaller as the particle loses energy and starts mirroring closer to the stellar surface. This panel also shows the PD and HR schemes taking $\sim 10$ larger time steps than the RKF scheme, meaning that these particles encounter many more mirrors with the same amount of iterations.  

\subsection{ODE Benchmarks} \label{sec:3.3}
To compare the utility of the adaptive schemes, we set up a benchmark scenario to assess the runtimes of each scheme directly. Our benchmark scenario is a static dipole $B$-field with surface $B$-field strength of $B_{\rm s}=10^{8} \, \rm{G}$, $\alpha = 0.0^\circ$, $R_{\rm s} = 0.01R_{\odot}$, $P_{\rm s} = 117$~s, $\gamma_{0} = 10^{4}$, and $\theta_{\rm p} = 160^{\circ}$. We initialise the particle at $0.15R_{\rm{LC}}$ and stop when the particle has been mirrored and has again reached $0.15R_{\rm{LC}}$. We ran the code with and without the RRF to assess the run times, and we were mainly interested in the relative energy error (final error after completion of the run) when the RRF was included. The reason we decided on this scenario as a benchmark is because it is similar to what we will be using when modelling AR Sco and where we will be producing emission maps and spectra for in future work. We normalised the run times to the time the RKF ran without an RRF for easy comparison. In Table~\ref{Benchmark1} in the second column, without the RRF, one sees that the higher-order schemes are significantly faster than the RKF, with the HR scheme being $\sim 20$ times faster. Column~3 shows that when including the RRF, the higher-order schemes are still much faster than the RKF scheme, but not as much as without the RRF, since the function evaluations become much longer when the RRF is included. With the RRF included, the HR is only $\sim 17$ times faster than the RKF. Evaluating the relative error in energy in column 4, we see that the schemes have similar small relative errors, but the higher-order schemes have slightly larger errors. Thus, it is important to balance runtime, stability, and accuracy when choosing a numerical scheme. 

To benchmark against the Vay scheme, we can not add the RRF or the adaptive methods. To address this, we use the same parameter setup as for Table~\ref{Benchmark1} but we specify a relative error threshold $\gamma_{\rm err}\sim10^{-10}$ and manually change the time step as a fraction of $P_{\rm g}$ for each scheme to achieve that error threshold. We can then see what fraction of $P_{\rm g}$ is required for each scheme to yield the specified error tolerance and assess the runtime of each scheme. Similar to Table~\ref{Benchmark1} we normalise the runtime of the schemes to the RKF runtime for easy comparison. These results are presented in Table~\ref{Benchmark2} where we see in column 3 that the higher-order schemes can take much larger time steps for the same accuracy since they need to sample $P_{\rm g}$ less, requiring smaller factions of $P_{\rm g}$. What is highlighted in the table is how well Vay performs in accuracy and runtime since, it has fewer function evaluations, outperforming the higher-order schemes in runtime even when using larger time steps. Unfortunately one has to manually run the simulation a few times to identify an applicable time step or in this case a scaling of $P_{\rm g}$ for each scenario. This becomes even more of a problem when introducing $E$-fields and RRFs that cause $P_{\rm g}$ to change rapidly. Beyond not being able to add the RRF, we will see in the next section that the Vay scheme begins to struggle when adding a large $E_{\perp}$-field as the problem becomes stiffer, requiring higher accuracy.    

\begin{table}
\begin{center}
\begin{tabular}{||c|c|c|c||} 
 \hline
  & Total runtime: & Total runtime: & Relative error \\ [0.5ex] 
 Method & No losses & RRF & $E_{\rm{err}}$ \\ [0.5ex] 
 \hline\hline
 RKF & 1.0 & 2.100 &  $4.771\times 10^{-3}$\\ 
 \hline
 DV & 0.4365 & 0.9304 & $4.767\times 10^{-3}$\\
 \hline
 PD & 0.0674 & 0.1473 & $4.829\times 10^{-3}$\\
 \hline
 CV & 0.0464 & 0.0789 & $4.934\times 10^{-3}$\\
 \hline
 HR & 0.0481 & 0.1202 & $4.922\times 10^{-3}$\\ [1ex] 
 \hline
\end{tabular}
\caption{\label{Benchmark1} Benchmark runtimes and relative error of the particle energy for a static dipole $B$-field without losses (column 2) and with the RRF included (column 3) for each numerical method indicated in column 1. The runtimes are normalised to the RKF method's runtime when there are no losses.}
\end{center}
\end{table}

\begin{table}
\begin{center}
\begin{tabular}{||c|c|c|c||} 
 \hline
 Method & Total runtime & Relative error $\gamma_{\rm err}$ & Time step: $P_{\rm g}$ fraction \\ [0.5ex] 
 \hline\hline
 RKF & 1.0 & $1.25\times 10^{-10}$ & $\frac{P_{\rm g}}{2000}$\\ 
 \hline
 DV & 0.487 & $3.89\times 10^{-10}$ & $\frac{P_{\rm g}}{900}$\\
 \hline
 PD & 0.03 & $1.96\times 10^{-10}$ & $\frac{P_{\rm g}}{35}$\\
 \hline
 CV & 0.038 & $2.43\times 10^{-10}$ & $\frac{P_{\rm g}}{35}$\\
 \hline
 HR & 0.047 & $1.85\times 10^{-10}$ & $\frac{P_{\rm g}}{20}$\\ [1ex] 
 \hline
 Vay & 0.016 & $2.20\times 10^{-10}$ & $\frac{P_{\rm g}}{50}$\\ [1ex] 
 \hline
\end{tabular}
\caption{\label{Benchmark2} Benchmark runtimes and relative error for a static dipole $B$-field without losses. The methods are given in column 1 with the runtimes normalised to the RKF method's runtime in column 2. Column 4 shows the time step scaling needed to obtain the required total relative $\gamma_{\rm err}$ in column 3.}
\end{center}
\end{table}

\subsection{Reproducing Takata et al.'s results} \label{sec:3.4}
\begin{figure}
\centering
\includegraphics[width=.5\textwidth]{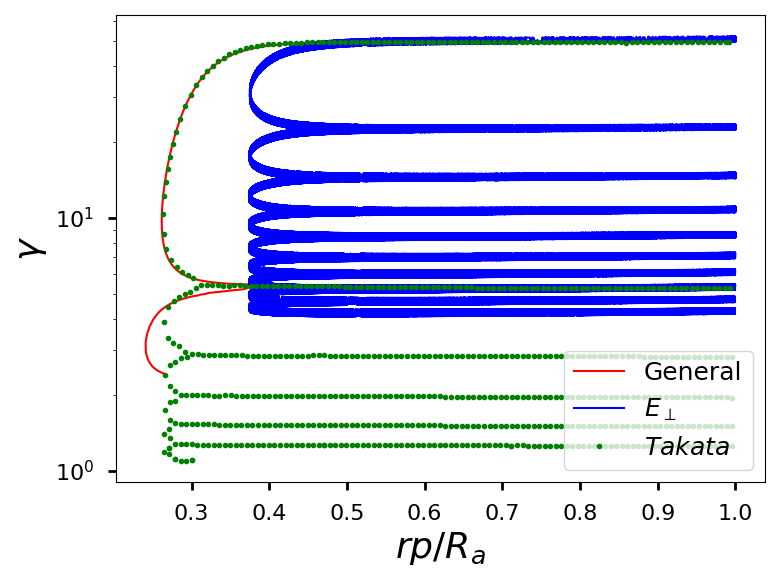}
\caption{Results from a magnetic mirror model proposed for AR Sco by \citet{Takata2017}. We used the same parameters as discussed in Section~\ref{sec:2.6}. The plot shows our results using the general RRF equations in red, our results when including an $E_{\perp}$ in blue, and the results from \citet{Takata2017} in green.}
\label{Takata_rep}
\end{figure}

In Section~\ref{sec:2.6}, we discussed the setup and parameters used to reproduce the magnetic mirror scenario for AR Sco as implemented in \citet{Takata2017}. In Figure~\ref{Takata_rep}, we plot the particle Lorentz factor $\gamma$ vs radial distance, normalised to the binary separation $R_{\rm a}$. We plotted our generalised RRF results for this scenario in red, the results from \citet{Takata2017} in green, and our model results when including an $E_{\perp}$-field defined by Equation~(\ref{E-Dipole}) in blue. We used the adaptive PD scheme for our model results without assuming super-relativistic particles with small pitch angles, in contrast to \citet{Takata2017}. We see that the general RRF results agree surprisingly well with those of \citet{Takata2017}, but we do see some deviation after the first magnetic mirror point. The reason for agreement could be due to how small the RRF is in this scenario, with the ratio of the RRF to the Lorentz force being $\sim 2\times10^{-11}$. The relative error in energy for our results was $E_{\rm err}=7.9\times 10^{-3}$. When including an $E_{\perp}$-field, we see that $\gamma$ decreases much slower, as the particle mirrors seven times before it has lost the same amount of energy as the particle at the first mirror in the previous case. The particle also mirrors farther away from the WD surface in this case. When including an $E_{\perp}$-field, $\gamma$ increases as the particle is moving in the same direction as the $E_{\perp}$-field, and then decreases as it moves opposite to the field. This creates an oscillatory effect seen in the blue curve. Unfortunately, we can not compare the particle pitch angle evolution, since \citet{Takata2017} do not show these specific results. We expect quite a big difference in the respective pitch angle evolution properties between our approaches, since Equations~(\ref{Takata_transport}) do not allow feedback due to the particle dynamics, namely drift, mirroring, and focusing effects. Moreover, those equations are valid in the limit of small pitch angles. We will study this aspect in more detail and insight in our next paper, since we will be able to compare our results in detail with the results of \citet{Harding2021} for non-uniform fields in a pulsar scenario.    

\subsection{Uniform $E$-field scenarios with no RRF} \label{sec:3.5}
As mentioned in Section~\ref{sec:2}, to test the accuracy and stability of our results when including an $E$-field, we calculate the transformed gyro-radius $(R_{\rm g}')$ and Lorentz factor $(\gamma ')$ in the frame co-moving with the particle $\mathbf{E}\times \mathbf{B}$-drift velocity. In the trajectory plots, we show the 2D $xy$-plane of the particle in the observer frame, as well as the frame co-moving with the particle $\mathbf{E}\times \mathbf{B}$-drift velocity. The particle position in the co-moving frame is normalised to $R_{\rm g}'$, but not in the observer frame, since $R_{\rm g}$ changes constantly due to the $\mathbf{E}\times\mathbf{B}$-drift effects. The relative error in Lorentz factor $\gamma_{\rm err}' = (\gamma_{i+1}' - \gamma_{i}')/\gamma_{i}'$ is plotted over time normalised to $P_{\rm g}'$ showing the numerical error in $\gamma$ per time step. The relative error in $R_{\rm g}'$ is the same as in the previous scenarios, but now using the transformed frame quantities and plotted over normalised time. In the transformed frame, we expect the particle to have a circular gyration in the 2D position plot (similar to the uniform $B$-field case) since there is no $E_{\perp}$ in the transformed frame to cause the particle to drift. Thus if the particle drifts in this frame, we expect this to be due to numerical instability.

\begin{figure*}
\centering
\includegraphics[width=\textwidth]{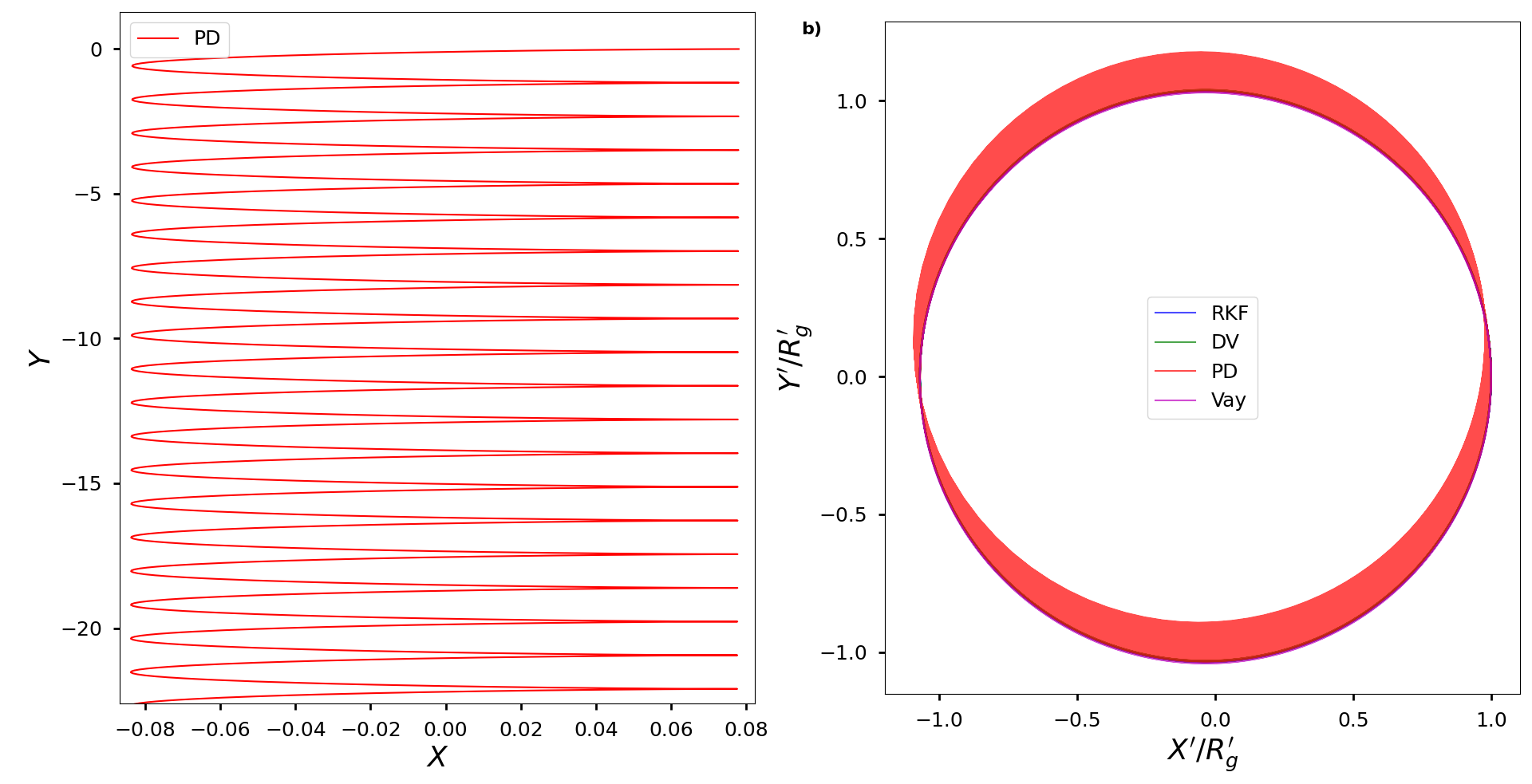}
\caption{Results for the $\mathbf{E}\times\mathbf{B}$-scenario using a constant $B$-field and $E_{\perp}$-field, and zero initial momentum. In panel a) we show the particle trajectory in the observer frame and in panel b) the particle trajectory in the transformed frame co-moving with the $\mathbf{E}\times\mathbf{B}$-drift velocity. For the Vay scheme, we used $dt = 10^{-14}$.}
\label{E_cross_B_P_zero_large_E_traj}
\end{figure*}

For the uniform $E_{\perp}$- and $B$-field scenario with lower fields, we use $B_z = 10^{5} \, \rm{G}$ and $E_x = 0.9B_z$. We set the initial momentum to zero ($\gamma_0=1$) to allow comparison with the scenario of \citet{Petri2017, Ripperda2018}, and initialised the particle to gyrate around the point (0,0,0) in the transformed frame. As a comparison, we included the Vay scheme using $dt = 10^{-13} \, \rm{s}$. In Figure~\ref{E_cross_B_P_zero_large_E_traj} panel~a) we see that the particle drifts in the negative $y$-direction in the observer frame, as expected. We see that at some points, $R_{\rm g}$ is small and that a small enough time step is necessary to sufficiently sample the gyro-period. We only plotted the PD scheme to show the trajectory, since panel b) is more useful for identifying stability. In panel b) we see that since there is no $E_{\perp}$ in the co-moving frame, the particle gyrates around the $B$-field, similar to the uniform $B$-field scenario in the lab frame. We have not plotted results from the CR and HR schemes, since they were found to drift, making it difficult to see all the results from all the schemes. The higher-order schemes seem to be more unstable for these extreme $E$-field scenarios, even though they have a higher accuracy. We believe this instability is due to the adaptive time step methods not using a sufficiently small time step for the higher-order schemes. The PD starts to drift a little after many gyrations. Notably, these instabilities were also found by \citet{Petri2017, Ripperda2018}. It is important to remember this is a high-field scenario with an extremely small $R_{\rm g}$, since the particle starts with no momentum, and the $E$-field strength is a high fraction of the assumed $B$-field. We found that the higher the $E$-field strength compared to that of the $B$-field, the more unstable the results become, similar to what was found by \citet{Ripperda2018}. This is due to the ODE becoming more stiff. 

\begin{figure*}
\centering
\includegraphics[width=\textwidth]{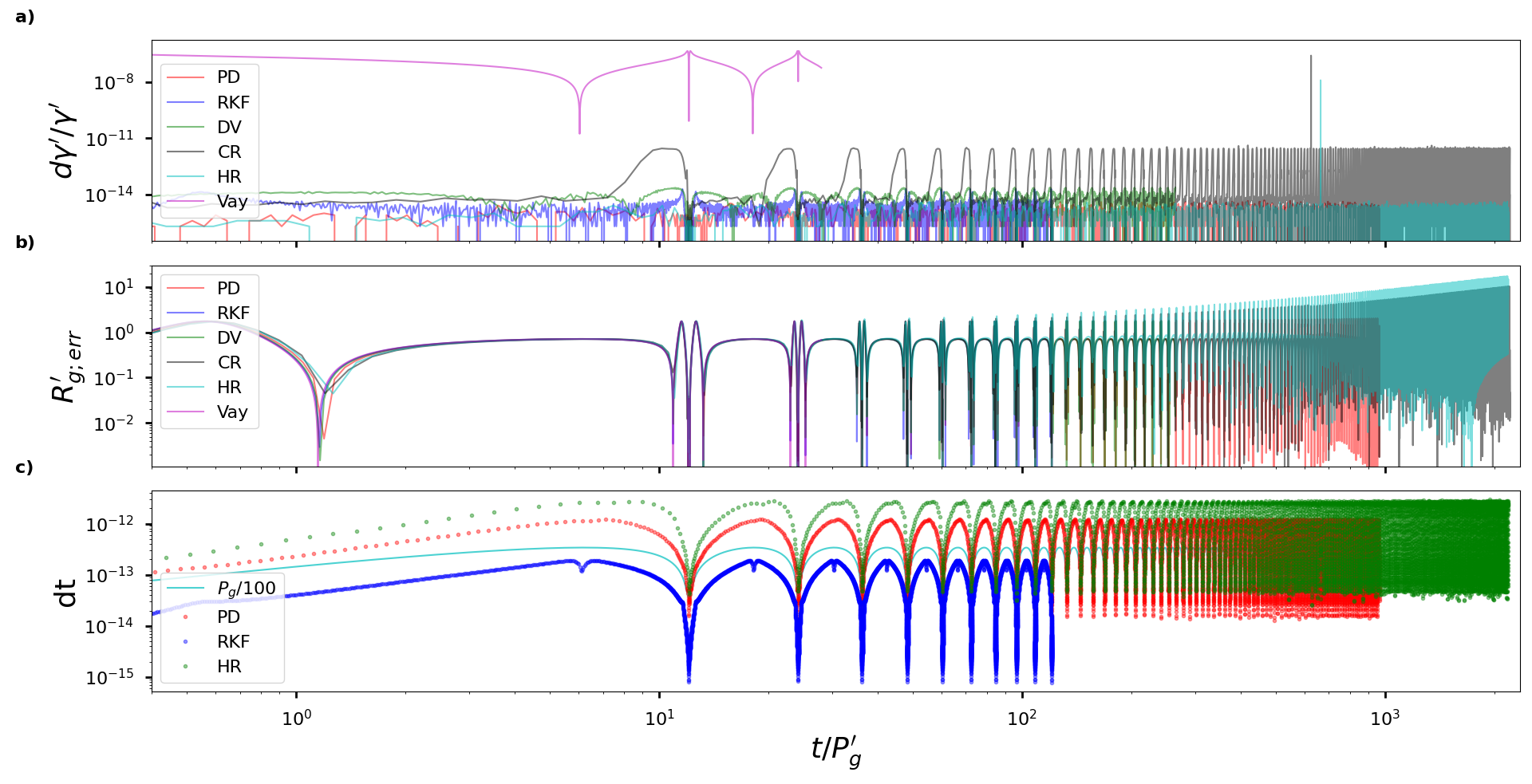}
\caption{Results for the $\mathbf{E}\times\mathbf{B}$-scenario using a constant $B$-field and $E_{\perp}$-field, and zero initial momentum. Panel a) shows the relative error in $\gamma '$ for each time step, panel b) shows the relative error in $R_{\rm g} '$, and panel c) shows the time step sampling of three methods. For the Vay scheme, we used $dt = 10^{-14}$.}
\label{E_cross_B_P_zero_large_E_err}
\end{figure*}

In Figure~\ref{E_cross_B_P_zero_large_E_err} panel a) we see that all the adaptive schemes are accurate, with relative errors $\gamma_{\rm err}'\sim 10^{-14}$, except for the CR scheme. Interestingly, the Vay scheme is found to have a high relative error of $\gamma_{\rm err}'\sim 10^{-7}$, so that each step would affect the seventh decimal and would cause significant numerical inaccuracies. Thus, using the Vay scheme in this scenario, one would only be able to do $\sim 10^{7}$ iterations before the numerical error is of the order of the results itself. When resolving the full particle gyration, one can easily go beyond $10^7$ iterations, especially if longer simulation time per particle is needed; or one would have to use extremely small time steps. The Vay method is still stable in this regime, as is seen in Figure~\ref{E_cross_B_P_zero_large_E_traj} and has a much more reliable accuracy when using a smaller $E$-field fraction without having to use significantly smaller time steps than the adaptive schemes. This highlights the importance of the high-order schemes in extreme cases to ensure both stability and accuracy. The relative errors from our results are of the same order of magnitude as those found in \citet{Petri2017, Ripperda2018}. In panel~b) of Figure~\ref{E_cross_B_P_zero_large_E_err} we note that the RKF, DV, and Vay schemes are quite stable without an increasing trend in their error. There is a small increasing trend visible in the PD scheme, but a noticeable increase in the CR and HR schemes. The relative error in $R_{\rm g}'$ is reasonably high at $R_{\rm g; err}'\sim 10^{0}$ but \citet{Ripperda2018} found similarly high errors in $R_{\rm g}'$ for a similar scenario. This could be because it is difficult to accurately centre the particle to gyrate around the origin (0,0,0) in the transformed frame. In panel c) we see that unlike in the uniform $B$-field scenario, in this scenario there is a considerable variation in the time steps used, where the schemes have to use $\sim 100$ times smaller time steps when the gyro-radius is small. In this scenario, we see that the higher-order schemes complete $\sim 10$ times more gyrations than the RKF scheme. What should be noted is how stable the RKF is: although it is slower and has lower-order accuracy, it is a good scheme to compare to. 

\begin{figure*}
\centering
\includegraphics[width=\textwidth]{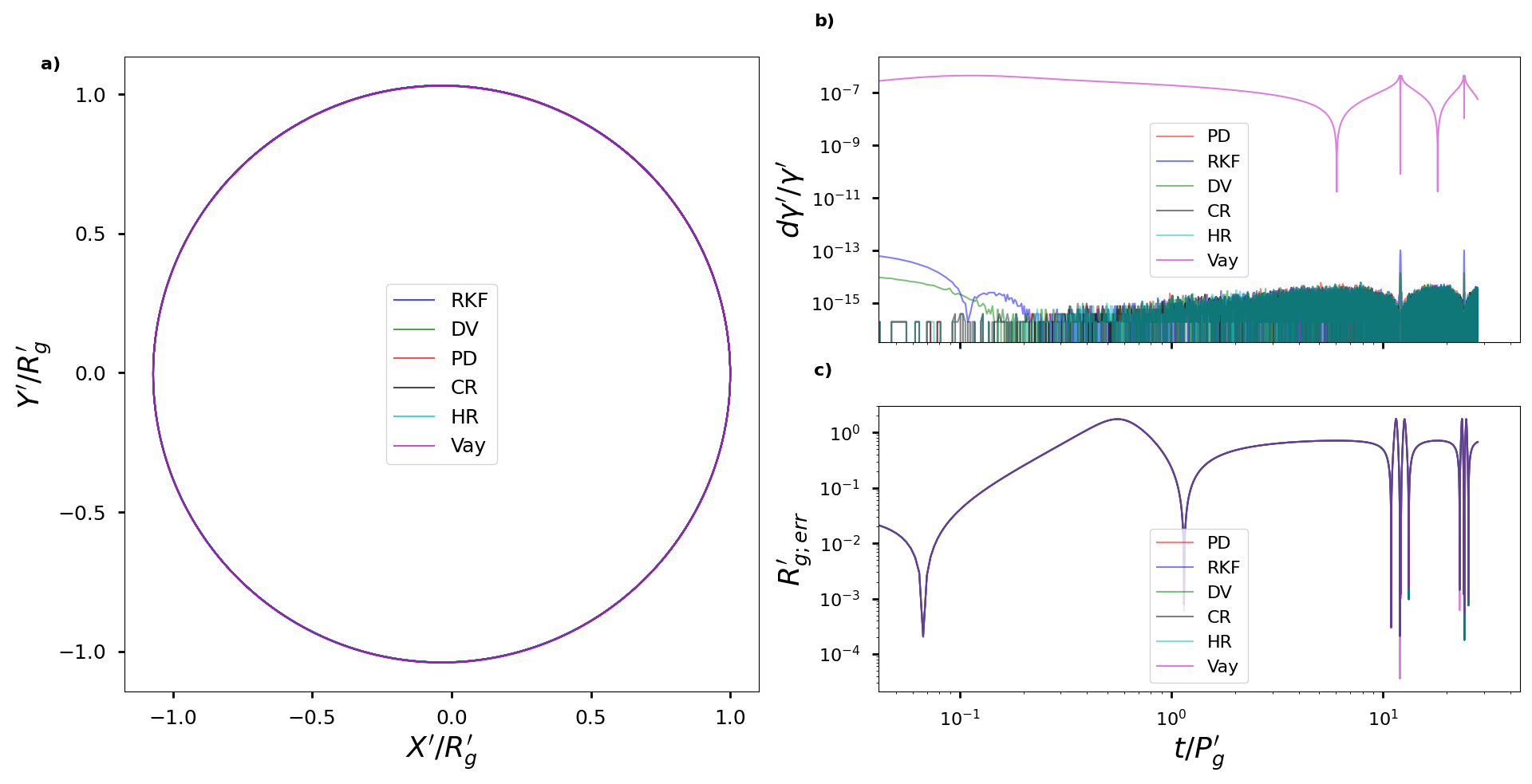}
\caption{Results for the $\mathbf{E}\times\mathbf{B}$-scenario using a constant $B$-field and $E_{\perp}$-field, with the same parameters as in Figures~\ref{E_cross_B_P_zero_large_E_traj} and~\ref{E_cross_B_P_zero_large_E_err}, but using a constant time step $dt=10^{-14}$.}
\label{E_cross_B_P_zero_large_E_const_dt}
\end{figure*}

For Figure~\ref{E_cross_B_P_zero_large_E_const_dt}, we used the same parameters as in the previous scenario, but invoked a constant time step $dt = 10^{-14} \, \rm{s}$ for each scheme. In panel a) we see that all the schemes are now stable, with no drifting observable. The only scheme that showed some instability if one zoomed in was the Vay scheme which was found to drift slightly. In panel b) it is more evident that when using the same time step for each scheme, the Vay scheme is significantly less accurate in this extreme regime to the point that one would worry about numerical inaccuracies. All schemes have relative errors of $\gamma_{\rm err}'\sim 10^{-15}$, with the Vay scheme at $\gamma_{\rm err}'\sim 10^{-8}$ (with max at $10^{-7}$). In panel c) we observe no increasing trend in the error, but with slightly lower relative errors than the previous adaptive time step results, with $R_{\rm g; err}'\sim 10^{-1}$. It must be noted that one may obtain a reasonable error for the Vay scheme, but then the time steps would have to be significantly smaller than what even the RKF requires to be stable and accurate.    

\begin{figure*}
\centering
\includegraphics[width=\textwidth]{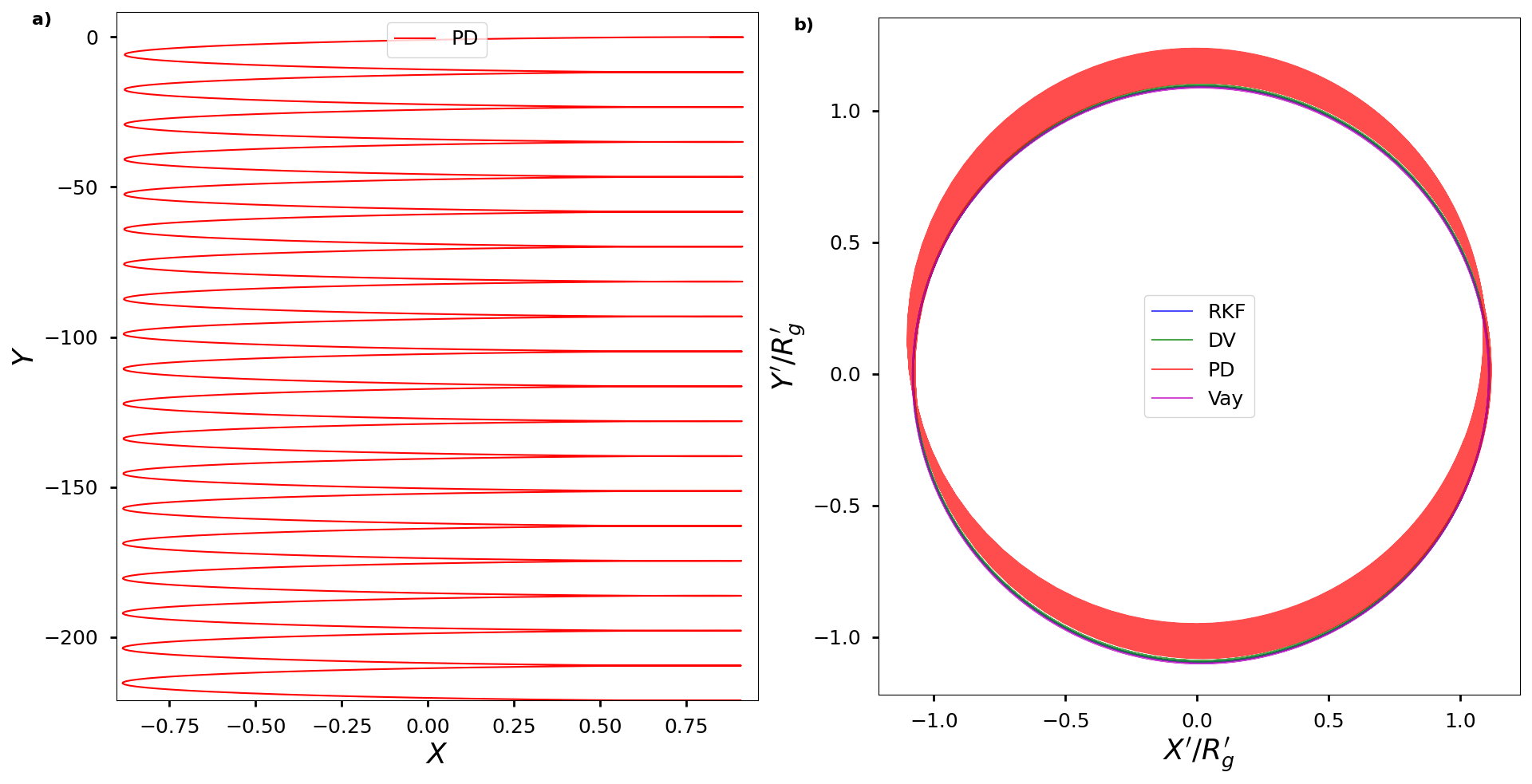}
\caption{Results for the $\mathbf{E}\times\mathbf{B}$-scenario for large fields, using a constant $B$-field and $E_{\perp}$-field and a large initial Lorentz factor. In panel a) we show the particle trajectory in the observer frame, and in panel b) the particle trajectory in the $xy$-plane in the frame co-moving with the $\mathbf{E}\times\mathbf{B}$-drift velocity. For the Vay scheme, we used a constant time step $dt = 10^{-13}$.}
\label{E_cross_B_large_E_gam_traj}
\end{figure*}

For the scenario where $E_{\perp}$- and $B$-field are uniform, but with a large Lorentz factor and higher fields, we use the fields $B_z = 10^{12} \, \rm{G}$ and $E_x = 0.9B$. We set the initial parameters to $\gamma_{0} = 10^{8}$, $\theta_{\rm p} = 90^{\circ}$, and use $dt = 10^{-13} \, \rm{s}$ for the Vay scheme. In Figure~\ref{E_cross_B_large_E_gam_traj} panel a) we see the particle drifts in the negative $y$-direction, with a small $R_{\rm g}$ due to the large $E$-field compared to the $B$-field. The larger the $E$-field compared to the $B$-field, the smaller the gyro-radius becomes at the turning point as the particle travels from parallel to $E_{\perp}$ to anti-parallel. In panel b) we note that the RKF, DV, and Vay schemes are quite stable whereas the PD appears reasonably stable with a minor drift observed over many orbits. The CR and HR schemes were once again not included, since they were found to drift, making it difficult to see all the schemes on one plot. 

\begin{figure*}
\centering
\includegraphics[width=\textwidth]{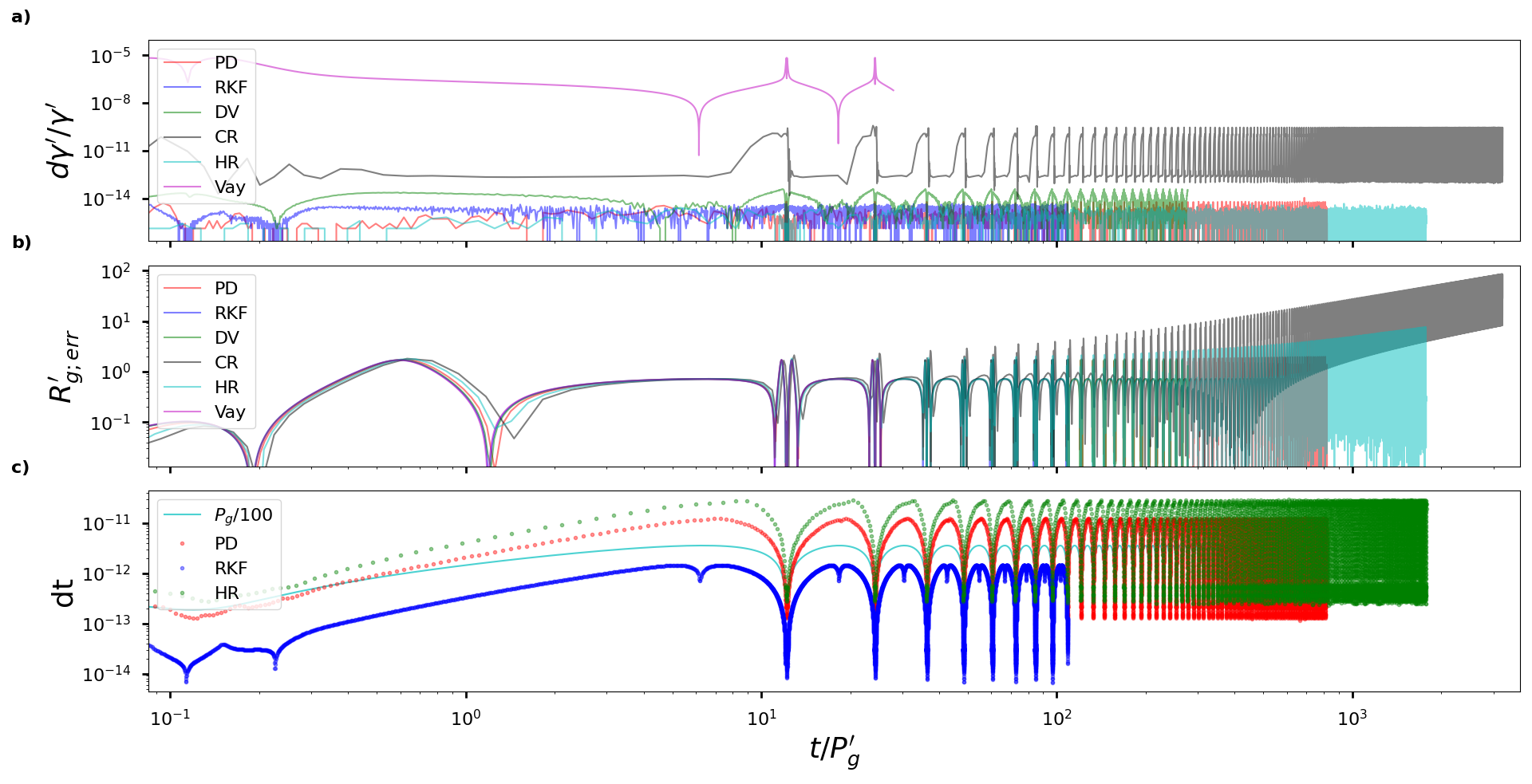}
\caption{Results for the $\mathbf{E}\times\mathbf{B}$-scenario for large fields, using constant $B$-field and $E_{\perp}$-field and a large initial Lorentz factor. Panel a) shows the relative error in $\gamma '$ at each time step, panel b) shows the relative error in $R_{\rm g}'$, and panel c) shows the time step sampling of three schemes. For the Vay scheme, we used a constant time step $dt = 10^{-13} \, \rm{s}$.}
\label{E_cross_B_large_E_gam_err}
\end{figure*}

In Figure~\ref{E_cross_B_large_E_gam_err} panel a) we see that all the schemes have good accuracy, even though the CR and HR schemes are a bit unstable. The errors of the schemes were found to be $\gamma_{\rm err}'\sim 10^{-14}$, except for the higher value of the CR scheme. The Vay scheme is once again found to be significantly less accurate in this extreme scenario, with a relative error of $\gamma_{\rm err}'\sim 10^{-6}$ (peaks at $10^{-5}$); thus, we would not recommend using the scheme in this scenario. In panel b) we observe an increasing error trend in the CR and HR schemes, with a small trend in the PD scheme, whereas the RKF, DV, and Vay schemes are quite stable. The stable schemes were found to have relative gyro-radius errors $R_{\rm g; err}'\sim 10^{-1}$ (peaks at $10^{0}$). 

\begin{figure*}
\centering
\includegraphics[width=\textwidth]{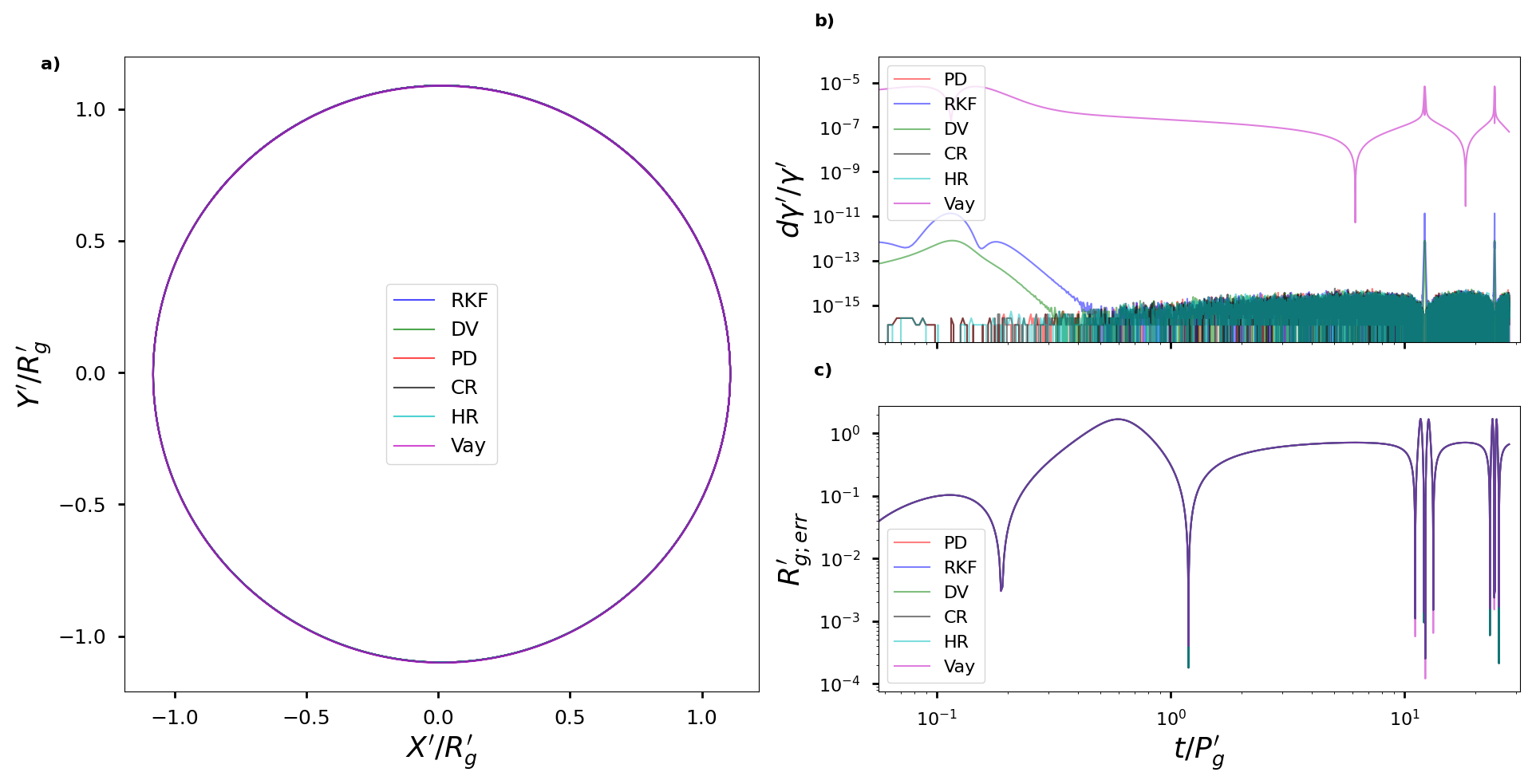}
\caption{Results for the $\mathbf{E}\times\mathbf{B}$-scenario for large fields, using a constant $B$-field and $E_{\perp}$-field with the same parameters as in Figure~\ref{E_cross_B_large_E_gam_traj} and~\ref{E_cross_B_large_E_gam_err}, but using a constant time step $dt=10^{-14} \, \rm{s}$.}
\label{E_cross_B_large_E_gam_const_dt}
\end{figure*}

Figure~\ref{E_cross_B_large_E_gam_const_dt} is for the same parameters as the previous scenario, but using a constant time step $dt = 10^{-13} \, \rm{s}$. In panel a) all the methods are once again stable, with no observable drift when using a constant time step below their stability threshold. In panel b) we see all the schemes are significantly more accurate than the Vay scheme when using the same constant time step. The adaptive schemes were found to have relative errors of $\gamma_{\rm err}'\sim 10^{-14}$, with the Vay scheme at $\gamma_{\rm err}'\sim 10^{-6}$. In panel c) there is no observable increase in the error trend for any of the schemes, with relative errors $R_{\rm g; err}'\sim 10^{-1}$.    

\subsection{$E$-field RRF scenarios} \label{sec:3.6}
\begin{figure*}
\centering
\includegraphics[width=\textwidth]{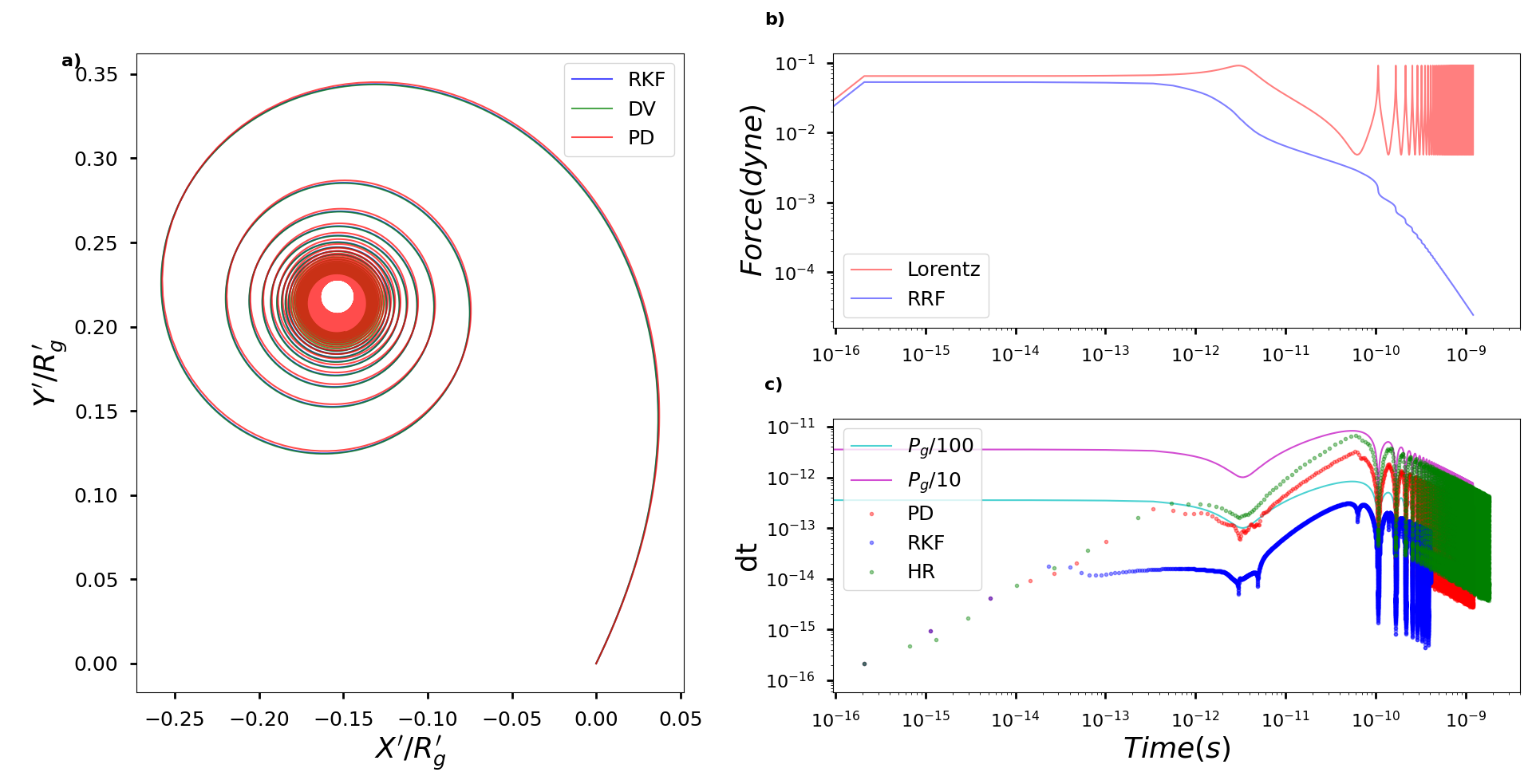}
\caption{Results for the $\mathbf{E}\times\mathbf{B}$-scenario with the RRF included, using a constant $B$-field and $E_{\perp}$-field. Panel a) shows the particle trajectory in the $xy$-plane of the frame co-moving with the $\mathbf{E}\times\mathbf{B}$-drift velocity. Panel b) shows the Lorentz force and the RRF, while panel c) shows the adaptive time step sampling for three of the schemes.}
\label{RRF_E_cross_B_large_E}
\end{figure*}

For the uniform $E_{\perp}$- and $B$-field with an included RRF, we use $B_z = 10^{8} \, \rm{G}$ and $E_x = 0.9B$. The initial parameters were $\gamma_{0} = 10^{4}$, $\theta_{\rm p} = 90^{\circ}$, and starting at the origin (0,0,0). In Figure~\ref{RRF_E_cross_B_large_E} panel a) we plot the 2D particle trajectory in the co-moving frame, with position components normalised to the initial $R_{\rm g}'$. Here we see that the schemes are reasonably well overlapping and stable, with no major drift observable. The CR and HR schemes do drift a little and were not included in the plot, for clarity. The particle $R_{\rm g}'$ decreases rapidly in this strong radiation-reaction scenario, spiralling inward as the particle loses energy. Panel~b) indicates the Lorentz force and the RRF vs time. The RRF is initially almost comparable in magnitude to the Lorentz force. The RRF then decreases, since there is no $E_{\parallel}$ to accelerate the particle to balance the RRF. The oscillations in the Lorentz force are due to the particle moving parallel to $E_{\perp}$ in one segment of the gyration and anti-parallel in the other. Panel c) indicates the time steps taken by the RKF, PD, and HR schemes over time. There is a large variation due to the changing $R_{\rm g}$ caused by the $\mathbf{E}\times \mathbf{B}$-drift. A general decrease is also seen due to the particle losing energy. We also note that the higher-order schemes are able to take much larger time steps than the RKF scheme.   

\begin{figure}
\centering
\includegraphics[width=.47\textwidth]{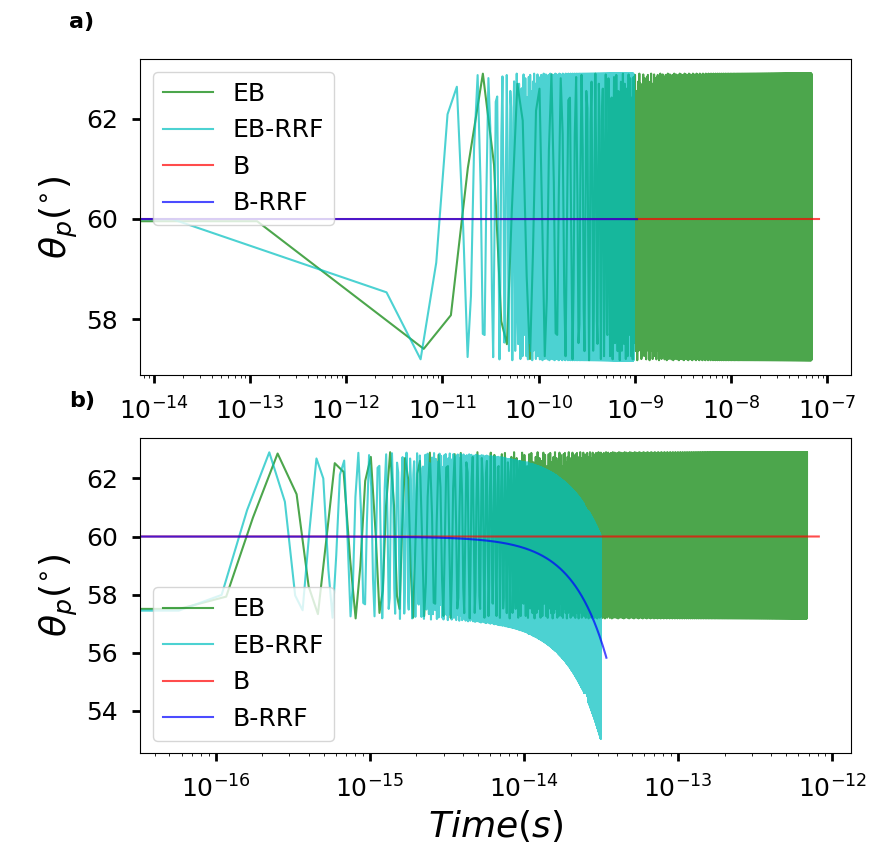}
\caption{Results for the pitch angle vs time for a uniform $B$-field in red, uniform $B$-field with RRF in blue, uniform $B$ and $E_{\perp}$-field in green, and uniform $B$ and $E_{\perp}$-field with RRF in cyan. In panel a) we used $B = 10^{8} \, \rm{G}$, $E_{\perp} = 0.1B$, and $\gamma=10^{4}$. In panel b) we use $B =10^{11} \, \rm{G}$, $E_{\perp} = 0.1B$, and $\gamma=10^{2}$.}
\label{Pitch_comp}
\end{figure}

Figure~\ref{Pitch_comp} shows the pitch angle evolution over time for the uniform field test scenarios with and without RRF. The scenarios include a uniform $B$-field (and zero $E$-field) without RRF plotted in red, and one with the RRF included plotted in blue. For the $\mathbf{E}\times \mathbf{B}$ scenarios, we use a uniform $B$-field and uniform $E_{\perp}$-field; the case without the RRF is plotted in green, and the one with the RRF is plotted in cyan. For all the cases, we used an initial pitch angle of $\theta_{\rm p}=60^{\circ}$. In panel a) we used $B_z = 10^{8} \, \rm{G}$, $E_{x} = 0.1B$, and $\gamma=10^{4}$. For panel b) we used $B_z =10^{11} \, \rm{G}$, $E_{\perp} = 0.1B$, and $\gamma=10^{2}$. In panel~a) we notice that for the $\mathbf{E}\times \mathbf{B}$ scenario, the pitch angle oscillates around the mean, namely the initial pitch angle. This is because in one gyration of the particle, the particle is moving parallel and anti-parallel with respect to the $E_{\perp}$-field, i.e., it is accelerated and decelerated by the $E_{\perp}$-field. This effect thus cancels out over one gyration so that the mean pitch angle stays constant. This holds even though the particle drifts due to the $E_{\perp}$-field, because the fields are uniform; as the particle drifts off one local $B$-field line, the pitch angle is the same on the next local $B$-field line. This does not hold for non-uniform fields, since the field strength and direction are different at various points. From panel a) we notice that for the $B$-field scenarios with and without the RRF, the pitch angle stays constant. In the $\mathbf{E}\times \mathbf{B}$ scenarios, both with and without the RRF, the mean pitch angle stays constant. In panel b) we observe that for the cases with the RRF, the pitch angle and mean pitch angle decrease. The reason the pitch angle is unaffected in panel a) is because in Equation~(\ref{Landau}), the third term is directly opposite to the particle momentum. This means the third term only scales the magnitude of the momentum vector and does not change the direction of the vector, which would change the particle pitch angle. Term 3 also scales as $\gamma^{2}$ compared to term 2; as the $\gamma$ increases, term 3 quickly becomes the dominant term in the RRF. Thus in panel a) the pitch angle seems constant, since $\gamma$ is high and term three dominates, but when $\gamma$ becomes small, the pitch angle decreases as seen in panel b) where we use a low $\gamma$ and high fields that term 2 becomes more relevant in the RRF expression. Importantly, one needs high $B$-fields or a low $\gamma$ for the RRF to be large enough to affect the particle velocity direction. The pitch angle will still decrease with low $B$-fields, but over a long time scale. That is why in panel a) the pitch angle does decrease due to term 2, but this decrease is exceptionally small and becomes more prominent at low $\gamma$. For super-relativistic particles, \citet{landau1975} defines the RRF to be opposite in direction of the velocity, as was similarly shown in \citet{Harding2006}, which is in agreement with our numerical results for the evolution of the pitch angle. In our next paper we will investigate the RRF's effect on the pitch angle in non-uniform fields since this has a significant effect on the synchrotron- and synchrocurvature radiation calculations. 

To compare our time step with the synchrotron cooling time, we calculate the latter for electrons, using $t_{\rm{cool}} = E/|dE/dt|$. Assuming an isotropic pitch angle, where $\langle \sin^2\theta \rangle = 2/3$, the mean radiated power becomes $\langle dE/dt \rangle = \sigma_{T}E^{2}B^{2}/6\pi m^{2}c^{3}$ for synchrotron radiation, where $\sigma_{T}$ is the Thomson cross section. For both panels a) and b) of Figure~\ref{Pitch_comp}, we found that our largest time step was 50 times smaller than $t_{\rm{cool}}$, thus we more than adequately sample for synchrotron radiation. To investigate whether the time it takes to reach a certain $\gamma$ during synchrotron radiation aligns with the theoretically expected time, we integrate $dE/dt = -kE^{2}B^{2}\sin^{2}\theta_{\rm p}$ where $k = \sigma_{T}/6\pi m^{2}c^{3}$. We use the initial section of the graph where the pitch angle is constant, namely $\theta_{\rm p} = 60^{\circ}$. After simplification, this yields $(\gamma_{0} - \gamma(t))/mc^{2}\gamma_{0}\gamma(t) = kB^{2}\sin^{2}\theta_{\rm p}(t_{0}-t)$. Using this formula, we found that for our results in panel a) we had a relative error in the time of $\sim 1.1\times10^{-3}$ and $\sim 4.4\times 10^{-3}$ for panel b). This is surprisingly close to what is expected, considering that we are using the general RRF and serves as another justification for the reliability of our results. When introducing the average pitch angle, i.e., $\langle\sin^{2}\theta_{\rm p}\rangle=2/3$, the relative error for panel~a) increases to $\sim 0.1$ and in panel~b) to $\sim 0.01$.        

\subsection{AE results} \label{sec:3.7}
The model of \citet{Harding2021} assumes that the super-relativistic particles moving in very large fields attain the AE limit that determines their trajectory. Thus, we use the following case to test a radiation-reaction limit scenario, and to assess if our results converge to the AE limit. We use a uniform $B$-field $B_z = 10^{8} \, \rm{G}$ and uniform $E$-field with perpendicular component $E_x = 10^{7} \rm{stat}\rm{V} \, \rm{cm}^{-1} = 0.1B$, and a parallel component $E_z = 10^{5} \rm{stat}\rm{V} \, \rm{cm}^{-1} = 0.001B$. We set the initial particle position at (0,0,0) with initial Lorentz factor $\gamma_{0} = 10^{6}$ and pitch angle $\theta_{\rm p} = 60^{\circ}$. For our model results, we used the adaptive PD scheme but still compared it with the other schemes as shown in Figure~\ref{AE_angle} panel a). 

\begin{figure}
\centering
\includegraphics[width=.47\textwidth]{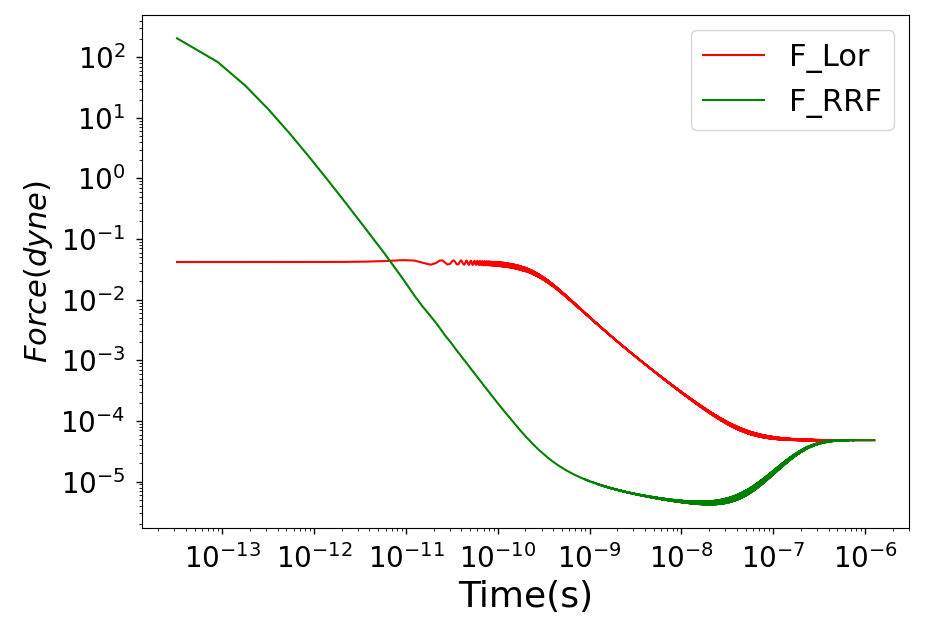}
\caption{Results for the radiation-reaction limit scenario, where the red line shows the Lorentz force and the green line the RRF when using our PD scheme.}
\label{AE_force_comp}
\end{figure}

\begin{figure}
\centering
\includegraphics[width=.45\textwidth]{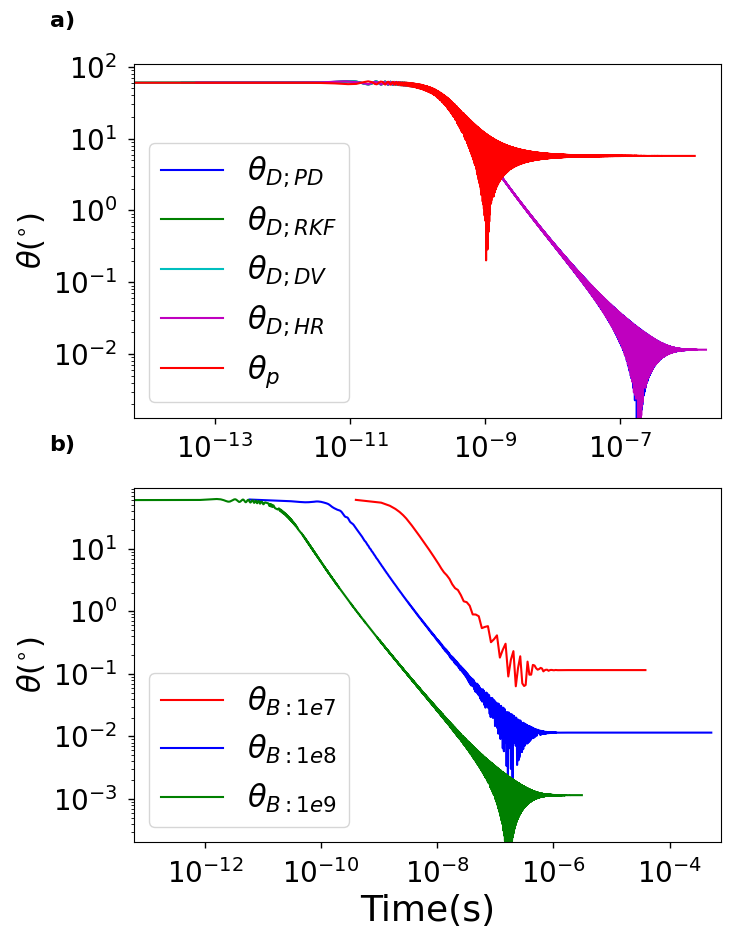}
\caption{Panel a) is the same scenario as in Figure~\ref{AE_force_comp} showing the angle between the model velocity and the AE velocity, where the pitch angle $\theta_{\rm p}$ is shown in red. Panel b) is the same scenario as in Figure~\ref{AE_force_comp}, but using different $B$-field strengths and $E_{\perp}$ components. In panel b) $E_{\perp} = 0.1\rm{B}$ is used in each case, while the red plot indicates $B=10^{7} \, \rm{G}$, blue $B=10^{8} \, \rm{G}$, and green $B=10^{9} \, \rm{G}$.} 
\label{AE_angle}
\end{figure}

In Figure~\ref{AE_force_comp}, we show the Lorentz force and the RRF for this scenario. We see that initially, the RRF is larger than the Lorentz force; we explained in Section~\ref{sec:2.4} that this can occur in the observer frame, one just needs to ensure that one does not violate the classical conditions for the RRF, which holds for our case (see Section~\ref{sec:2.4}). Initially, the RRF is high and drops rapidly in a short time, since the RRF's leading term has a $\gamma^{2}$ dependence. Thus, as $\gamma$ decreases so does the RRF. The accelerating electric field $E_{0}$ then accelerates the particle, increasing the RRF, until the losses and acceleration reach an equilibrium and the RRF and Lorentz forces become equal after a long time period. 

\begin{figure*}
\centering
\includegraphics[width=\textwidth]{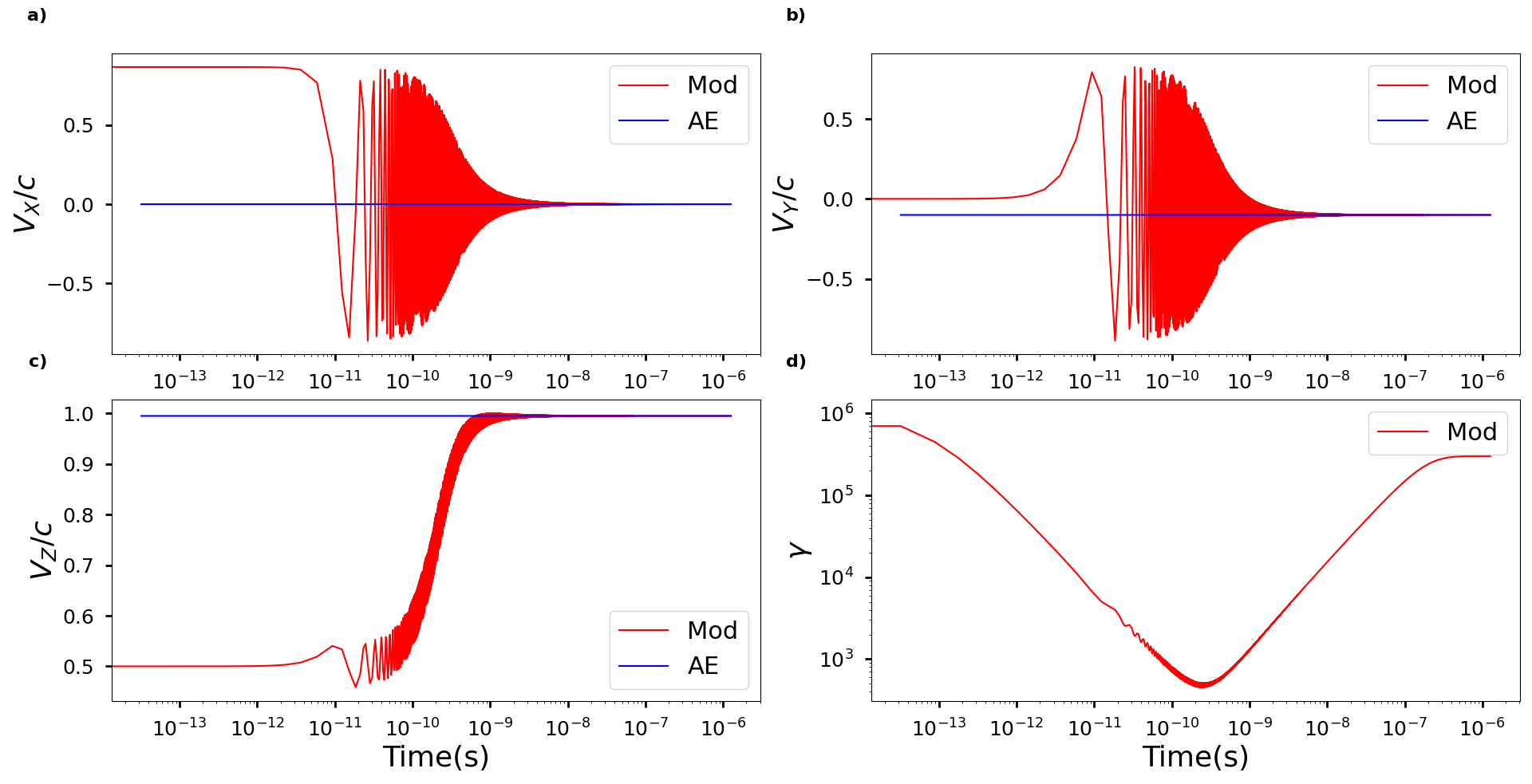}
\caption{The same scenario as in Figure~\ref{AE_force_comp}, where the red line represents our model results and the blue line the AE limit from Equation~(\ref{AE}). Panel a) - c) show each velocity component normalised to $c$. Panel d) shows the Lorentz factor, where the AE results are for Equation~(\ref{gam_crit}).}
\label{AE_vel_plt}
\end{figure*}

In Figure~\ref{AE_vel_plt}, we study the same scenario as before, but in panels a), b) and c) we plot each velocity component normalised to $c$. Our results are shown in red and the AE results are in blue. The AE velocity shown is the local AE velocity at each point. Our results initially deviate from the AE results, but eventually converge to the AE velocity components. Thus our results converge to the local AE velocity after some period and settle into the equilibrium state. In panel d) we plot the Lorentz factor from our model where we see a rapid decrease of $\gamma$ from our model due to the RRF and then an increase due to $E_{0}$, until it reaches an equilibrium where $\gamma$ becomes constant.

In Figure~\ref{AE_angle} panel a) we show the deviation angle $\theta_{D}$ between our predicted velocity using the different schemes and the AE velocity vs time. The red line is the standard pitch angle from our model results using the PD scheme for comparison. The pitch angle becomes smaller until it reaches a constant value at equilibrium. Importantly, the pitch angle does not become zero, because as $E_{\parallel}$ is accelerating the particle to high $\gamma$ values, the RRF does not decrease the pitch angle significantly as explained in the previous section. All the schemes give similar results for $\theta_{D}$. These results show $\theta_{D}$ becoming small and at the equilibrium, reaching $\sim 0.01^{\circ}$ but not zero. This is because Equation~(\ref{AE}) is gyro-centric and our result resolves the full particle gyration. This means that there is an offset from the gyro-centre in our results due to the particles gyro-radius and the non-zero pitch angle. Unfortunately, we have not found a reasonable and accurate method to obtain the gyro-centric velocity for a highly relativistic scenario including $E$-fields. The fact that at equilibrium we get a constant deviation angle means our particle is gyrating around the AE trajectory curve with a constant pitch angle, so that the AE direction coincides with our particle gyro-centre's direction of motion. This is further emphasised in panel b) where we show $\theta_{D}$ using the same scenario as before, but changing the uniform $B$-field strength, such that the gyro-radius at equilibrium is smaller for higher field values. These results were produced using the PD scheme, where the red curve is for $B_z = 10^{7} \, \rm{G}$, the blue curve for $B_z = 10^{8} \, \rm{G}$, and the green curve indicating $B_z = 10^{9} \, \rm{G}$. In each case, we also use the same $E_{\perp}$-field fraction for consistency in the particle $\mathbf{E}\times \mathbf{B}$-drift scenario, namely $E_x = 10^{7}\rm{stat}\rm{V} \, \rm{cm}^{-1}$. Here we see deviations for the red curve of $\sim 0.1^{\circ}$, for the blue curve of $\sim 0.01^{\circ}$, and for the green curve of $\sim 0.001^{\circ}$. Therefore, for higher uniform $B$-field values and smaller gyro-radii at equilibrium, the constant $\theta_{D}$ at equilibrium becomes smaller. This supports the conclusion that our results follow the AE limit as gyro-centre, and that the particle is gyrating around the AE trajectory curve at equilibrium. We have thus demonstrated that we recovered the AE limit via our numerical calculations.     

\section{Conclusion}\label{sec:4}
In preparation for first-principles modelling of the light curves and spectra of the white dwarf system AR Sco, we implemented general particle dynamics. We explained in detail the numerical methods we used to obtain our modelling results in order to ensure that they are reliable, clear, and reproducible. In all the test scenarios, we obtained the expected analytical results with good stability and accuracy, thus the numerical errors do not affect the results. We believe the results have important implications both for our application to AR Sco, where RRFs are high and magnetic mirroring changes the particle velocity rapidly, and for PIC simulations of pulsar magnetospheres. 

For the $B$-field scenarios, we found that, as expected, the high-order schemes gave better accuracy; but when adding a large $E$-field and the RRF, the lower-order schemes performed better due to them being more stable. The results do highlight that integration schemes can be accurate but have poor stability and vice versa, thus ensuring that both aspects are adequate is crucial for our astrophysical simulations. In our results, we assess our adaptive schemes without setting a limit to the maximum time step each scheme uses. We do this because unless one does prior runs for each scenario or initialisation configuration, one does not know what a sufficiently small time step is for each scheme. This is one of the major problems when using constant time steps or manually scaling the time steps as is used in implicit and symplectic schemes. 

We also show that our adaptive schemes are sufficiently stable and accurate for scenarios where the gyro-period drastically changes, such as in pulsar magnetospheres where the $B$-field strength rapidly drops with distance above the stellar surface. This ensures sufficient sampling within one gyration of the particle, even if the gyro-period is rapidly changing. The adaptive schemes can therefore take large time steps where there is a large gyro-period, saving enormous amounts of computational time and still maintaining accuracy and stability. This is currently a big problem in PIC codes that mainly use second-order accuracy symplectic integrators with constant time steps. This means that these codes have to start far away from the surface of the star or use unrealistically small magnetic fields to be able to use time steps that are large enough for the code to have a realistic runtime. We have also shown that the Vay scheme is unreliable in scenarios with high $E$-fields compared to the $B$-fields, because it is a second-order accuracy scheme. Thus one has to use exceedingly small time steps to ensure the results are accurate causing it to have an unfeasible runtime. 

Across all the test scenarios, we see that the adaptive time step method does introduce some instability in the higher-order schemes. This is because the unconstrained adaptive method we have implemented takes time steps that are slightly too large for the high-order schemes, causing instability or inaccuracy. We show that even though there is instability introduced in the schemes, the DV and PD schemes were both sufficiently accurate and stable in all cases, with the RKF being the most stable. As mentioned, in the methods, we used higher accuracy adaptive time step methods than is standardly used, as is required for realistic results in our pulsar modelling. Should better methods be found in future, these will be implemented to hopefully make the HR scheme more stable in the extreme scenarios, since it can take much larger time steps, saving significant computational time. Notably, by implementing an upper limit to the time steps, the HR scheme is stable and accurate, but we would have to do prior runs to find a sufficiently small upper limit without limiting the run time too much, suffering a similar problem as the constant-time-step solvers. When balancing run time, accuracy, and stability, we found the PD to be ideal for our simulations, since it is significantly faster and more accurate than the RKF scheme, but does not suffer too much instability due to the adaptive time step method as the CR and HR schemes do.

Our results for AR Sco connected to the magnetic mirror scenario of \citet{Takata2017} show agreement with theirs when using the same static vacuum dipole $B$-field. However, when introducing an $E_{\perp}$-field, the results differ drastically. We agree that $E_{\parallel}$ could be screened, but it is crucial to include $E_{\perp}$, since the $\mathbf{E}\times \mathbf{B}$-drift has a significant effect on the particle trajectories. This should have a big impact on the model light curves and spectra. It would have been ideal to compare the particle pitch angle evolution between our respective approaches. Since the transport Equations~(\ref{Takata_transport}) include no feedback on the particles and vice versa, one would have to manually turn the particle around at the magnetic mirror points. In future work, we will show full emission modelling for this magnetic mirror scenario of AR Sco with an added $E_{\perp}$-field.   

We lastly demonstrate in our results that even in these extreme scenarios with an added RRF, as occurs in pulsar magnetospheres, we find accuracy in our self-consistent energy conservation calculations. We furthermore obtain the AE results in the radiation-reaction limit. Importantly, we show the convergence to the AE results in the scenarios that we studied, which we have not seen done by other authors as of the writing of this paper. This is crucial, since the AE describes a solution to the equilibrium state, but not how a particle enters this equilibrium. There is thus ambiguity on how and when the particle enters this equilibrium to be able to use the AE equations to describe the particle motion. The closest model showing this transition into equilibrium is that of \citet{Yangyang2023} who discretised the problem into three regimes, but without consistently using the same equation of motion. In our next paper, we will show and analyse the AE comparison results for retarded dipole fields and force-free fields, with application to pulsar modelling, since this will have important implications on the radiation calculations required for the emission modelling. 

In conclusion, our results show that our particle dynamics calculations are sufficiently accurate, stable, and have significant runtime improvements. In our next paper, we will discuss all our emission modelling and compare our results to the pulsar emission code of \citet{Harding2021}. By doing this, we can show confidence in our emission modelling results as well as discuss the implication of comparing gyro-centric particle trajectories with full gyro-resolved particle dynamics. We will also discuss the applicability of the other assumptions used by this model and similar pulsar models. These assumptions include super-relativistic particles with small pitch angles and no feedback on the particle trajectories from the transport equations and vice versa. 

Due to our general approach to the particle dynamics, we believe our new code and methods we used can be applied to other sources as well (beside pulsars) where full particle dynamics are required. In the future, we will self-consistently calculate the field values, similar to other PIC codes. However for our next paper, we will use the force-free field grids for global pulsar magnetospheres that were obtained using a dissipative model solving the Maxwell equations. We also advise the use of the higher-order solvers with adaptive time steps in such cases, since we have shown that this improves both run time and the accuracy of the results.

\section*{Acknowledgements}
We thank the anonymous referee for comments that helped us improve the paper's presentation. This work is based on research supported wholly / in part by the National Research Foundation of South Africa (NRF). The Grant holder acknowledges that opinions, findings and conclusions or recommendations expressed in any publication generated by the NRF-supported research is that of the author(s), and that the NRF accepts no liability whatsoever in this regard.
L.~D. acknowledges partial support by the NRF under award numbers PMDS22060820024 and PMDS23042496534 as well as Anu Kundu for helpful discussion. C.~V. is funded by the research programme ``New Insights into Astrophysical and Cosmology with Theoretical Models Confronting Observational Data'' of the National Institute for Theoretical and Computational Sciences (NITheCS) of South Africa. Z.~W. acknowledges support by NASA under award number 80GSFC21M0002.

\section*{Data Availability}
The data underlying this article will be shared on reasonable request to the corresponding author.


\bibliographystyle{mnras}
\bibliography{ARsco_refs} 

\bsp	
\label{lastpage}
\end{document}